\documentclass[final,5p,times,twocolumn,numbers]{elsarticle}

\usepackage[utf8]{inputenc}
\usepackage[T1]{fontenc}
\usepackage{amsmath,amssymb,amsthm}
\usepackage{graphicx}
\usepackage{booktabs}
\usepackage{url}
\usepackage{natbib}
\usepackage{hyperref}
\usepackage{xcolor}
\usepackage{cleveref}
\usepackage{tabularx}
\usepackage{enumitem}
\usepackage{wrapfig}
\usepackage{array}
\hypersetup{
    colorlinks=true,
    linkcolor=blue,
    citecolor=black,
    urlcolor=blue
}

\usepackage{pgfplotstable}
\pgfplotsset{compat=1.18}

\newcolumntype{L}[1]{>{\raggedright\arraybackslash}p{#1}}
\usepackage{fancyhdr}
\usepackage{lastpage}

\theoremstyle{definition}
\newtheorem{definition}{Definition}
\begin{document}

\begin{frontmatter}

\title{Data Architectures and their Technical Requirements (DATER)}

\author[hsnr,rwth]{Sayed Hoseini\corref{cor1}}
\ead{sayed.hoseini@hsnr.de}
\author[hsnr,fit]{Christoph Quix}
\author[fit,rwth]{Stefan Decker}
\cortext[cor1]{Corresponding author.}

\address[hsnr]{Niederrhein University of Applied Sciences, Krefeld, Germany}
\address[fit]{Fraunhofer FIT, St.~Augustin, Germany}
\address[rwth]{RWTH Aachen University, Aachen, Germany}

\begin{abstract}
Modern organizations generate and consume massive volumes of heterogeneous data at high speed. This requires a continuous development of new techniques for more efficient and reliable data management. Designing appropriate data architectures has therefore become a strategic necessity, as they shape how data is integrated, governed, and made available for analytics and decision-making. 
This paper introduces a conceptual framework—\textbf{D}ata \textbf{A}rchitectures and their \textbf{Te}chnical \textbf{R}equirements (DATER)—to systematically describe and evaluate data architectures based on technical requirements. 
Six modern architectures are examined: data warehouse, (semantic) data lake, data lakehouse, data fabric, and data mesh. Each is analyzed by historical context, defining features, and conformance to DATER dimensions. The study supports researchers and practitioners in navigating architectural paradigms, clarifying overlaps, and highlighting strengths, limitations, and use-case suitability.
\end{abstract}


\end{frontmatter}

\section{Introduction}\label{chapter:data_architectures}
The exponential increase in the volume, velocity, and variety of data, fueled by digitalization, the Internet of Things, and Artificial Intelligence (AI), has made traditional, monolithic data systems insufficient. 
Traditional architectures such as data lakes and warehouses are increasingly complemented or replaced by emerging paradigms, because modern data ecosystems face growing demands for scalability, advanced analytics, interoperability, and decentralized governance. 
In this context, understanding and selecting the appropriate data architecture becomes a strategic imperative, as it directly affects the ability to derive value from data assets \cite{tzanetos2024introduction}.

In recent years, the landscape of data management has undergone a profound transformation and data architectures play a fundamental role. 
In information technology, a data architecture describes various aspects of system design in order to provide guidance for the development of data management system \cite{muller2008reference}. 
The common goal is the effective management of data and the various systems or components in which it resides.
Data architectures are most effective when they support the entire organization, enabling consistent data standardization and integration across all data-related activities \cite{groger2021there}.
Data architecture designs may thus vary for different levels within an organization, for particular domains, or based on its focuses such as infrastructure, applications, or data. 
 
This paper gives an overview of the current predominant data architectures. Choosing the appropriate data architecture is critical, as substantial changes in enterprise data management typically require significant organizational effort.
Gaining insights into the experiences, strategies, implementations, and deployment processes involved in transitioning to a modern architecture helps mitigate costly pitfalls and prevents data mess.
We focus on the technical characteristics for particular data architectures including their historical development and then discuss main differences and similarities. 
Our descriptions of the different architectures are intended to support the development of reference architectures \cite{10.1145/96602.96604,7916249,sang2017simplifying}. 
These reference architectures serve as blueprints for designing concrete solutions based on specific architecture types, offering standardized design guidelines \cite{van2003data} and fostering a shared vocabulary.

Furthermore, we propose a comprehensive framework for evaluating and classifying various data architectures referred to as \textbf{D}ata \textbf{A}rchitectures and their \textbf{T}echnical \textbf{R}equirements (DATER).
DATER provides valuable insights into the evolving landscape of data systems, helping both researchers and practitioners make informed decisions about which architectural models best suit their specific needs and goals.
While many real-world systems do not fully align with theoretical models, DATER enables a precise identification of which architectural components or capabilities differ. 
This helps bridge the gap between abstract design principles and practical implementations, guiding more effective system development and evaluation.

\subsection{Review methodology and Contributions}
This article follows a structured review methodology commonly applied in survey and review studies
\cite{badampudi2015experiences}. The review process is organized into three main stages:
(i) planning the review by defining the scope and research question,
(ii) conducting the literature search and screening process, and
(iii) reporting and synthesizing the findings.

\subsubsection{Scope and research question}
This review focuses on modern data architectures and architectural frameworks proposed in both academic research and industry practice.
The central research question motivating this review is:
\begin{description}
    \item \textit{What constitutes a structured framework for characterizing and differentiating modern data architectures, and how do existing architectures align with the technical requirements articulated by this framework?}
\end{description}

The scope is limited to data management within a single organization or enterprise context. Architectures primarily concerned with inter-organizational data sharing, such as data spaces \cite{otto2022evolution}, are excluded, as their primary focus lies on cross-organizational governance and collaboration rather than internal data architecture design. In particular, Gessler et al. \cite{gessler2025business} situate data spaces at the ecosystem level, not at the architecture level (see \cref{section:terminology}).

Related architectural concepts that address specific aspects of data processing, such as Lambda and Kappa architectures \cite{lin2017lambda}, or extensions of classical data warehouses (e.g., Data Vaults \cite{linstedt2015building,inmon2014data}), are considered only insofar as they contribute to the broader architectural landscape.

\subsubsection{Search criteria}
To ensure relevance and quality, the literature selection process was guided by explicit inclusion and exclusion criteria.

\paragraph{Inclusion criteria}
Publications were included if they met the following conditions:
\begin{itemize}[leftmargin=15pt]
    \item Publication in English and full-text availability through digital libraries or publisher platforms;
    \item Relevance to data architecture, data management architectures, or architectural frameworks;
    \item Coverage of concepts such as data warehouses, data lakes, lakehouses, data fabrics,  or comparable architectural paradigms;
    \item Recognized relevance in academic or industrial discourse, indicated by citation counts
    or adoption in practice.
\end{itemize}

\paragraph{Exclusion criteria}
Publications were excluded if they:
\begin{itemize}[leftmargin=15pt]
    \item Focused primarily on applications rather than reference architectures or data management;
    \item Addressed specialized processing architectures (e.g., stream-processing-only systems) without architectural considerations for data organization and governance;
    \item Concentrated on cross-organizational data ecosystems, such as data spaces or data marketplaces, without a focus on internal enterprise architectures.
\end{itemize}

Although a wide range of additional architectural terms exists, such as data hubs, grids,
federated or hybrid architectures, and event-driven or knowledge-graph-based architectures
\cite{linstedt2015building,7364082,lin2017lambda,hoseini2023sedar,plattform_industrie_2019_gaiax,venugopal2006taxonomy}, many of these can be traced back to a small set of core architectural patterns
\cite{ataei2022state}. Accordingly, this review concentrates on those architectures that have gained
substantial traction across both academia and industry.

\subsubsection{Search method}
The literature search was conducted primarily using Google Scholar, which forwarded to articles on  IEEE Xplore, ACM Digital Library, ScienceDirect, Springer, and related publisher platforms.
For each architectural paradigm, an initial set of approximately 20 well-cited journal articles or books was identified.

To extend and refine the literature set, citation snowballing was applied \cite{badampudi2015experiences}, including both backward searches through reference lists and forward searches using, for example, the ``Cited by'' functionality in Google Scholar.

In addition to academic literature, real-world systems referenced in the evaluation were identified using targeted web searches, technical documentation, and publicly available repositories on platforms such as GitHub. 

\subsubsection{Reporting and synthesis}
The results of the review are reported by synthesizing architectural concepts into a unified framework termed DATER. The framework enables a structured comparison of architectures based on technical requirements and evaluation dimensions, and serves as the basis for mapping real-world systems to architectural models in subsequent sections. The screened literature and synthesized technical requirements form the basis for deriving the evaluation criteria presented in \cref{section:evaluation}.

\paragraph{Selection of evaluated systems}\label{section:selection_of_systems}
The systems included in the evaluation were selected as representative implementations of the architectural paradigms discussed in this review rather than as an exhaustive survey of available platforms. In addition to well-established reference systems, we deliberately included systems whose architectural classification is ambiguous or hybrid in nature.

This choice allows us to demonstrate the analytical value of DATER in cases where existing architectural labels are insufficient or contested. By applying DATER to such systems, it becomes possible to pinpoint which architectural layers and technical requirements are fulfilled, partially addressed, or blurred across paradigms.

The number of evaluated systems was therefore intentionally limited, as additional systems with
similar architectural characteristics would not provide further insight beyond implementation-specific variations.

\paragraph{Contributions} In summary, this article makes the following contributions:
\begin{enumerate}[leftmargin=10pt]
    \item We first define our terminology to clarify key terms such as data ecosystem, data architecture, and data platform (\cref{section:terminology}). 
    \item The core of the DATER framework is the characterization, definition, and identification of technical requirements for six data architectures (\cref{section:data_architectures}).
    \item From this, we extract nine evaluation dimensions to assess data architectures. Moreover, a comparative evaluation of real-world systems is conducted by mapping them against the DATER framework, providing insights into their alignment with various architectural models (\cref{section:evaluation}).
\end{enumerate}

\section{Terminology}\label{section:terminology}
\begin{figure*}[tb]
		\centering
		\includegraphics[width=0.5\textwidth]{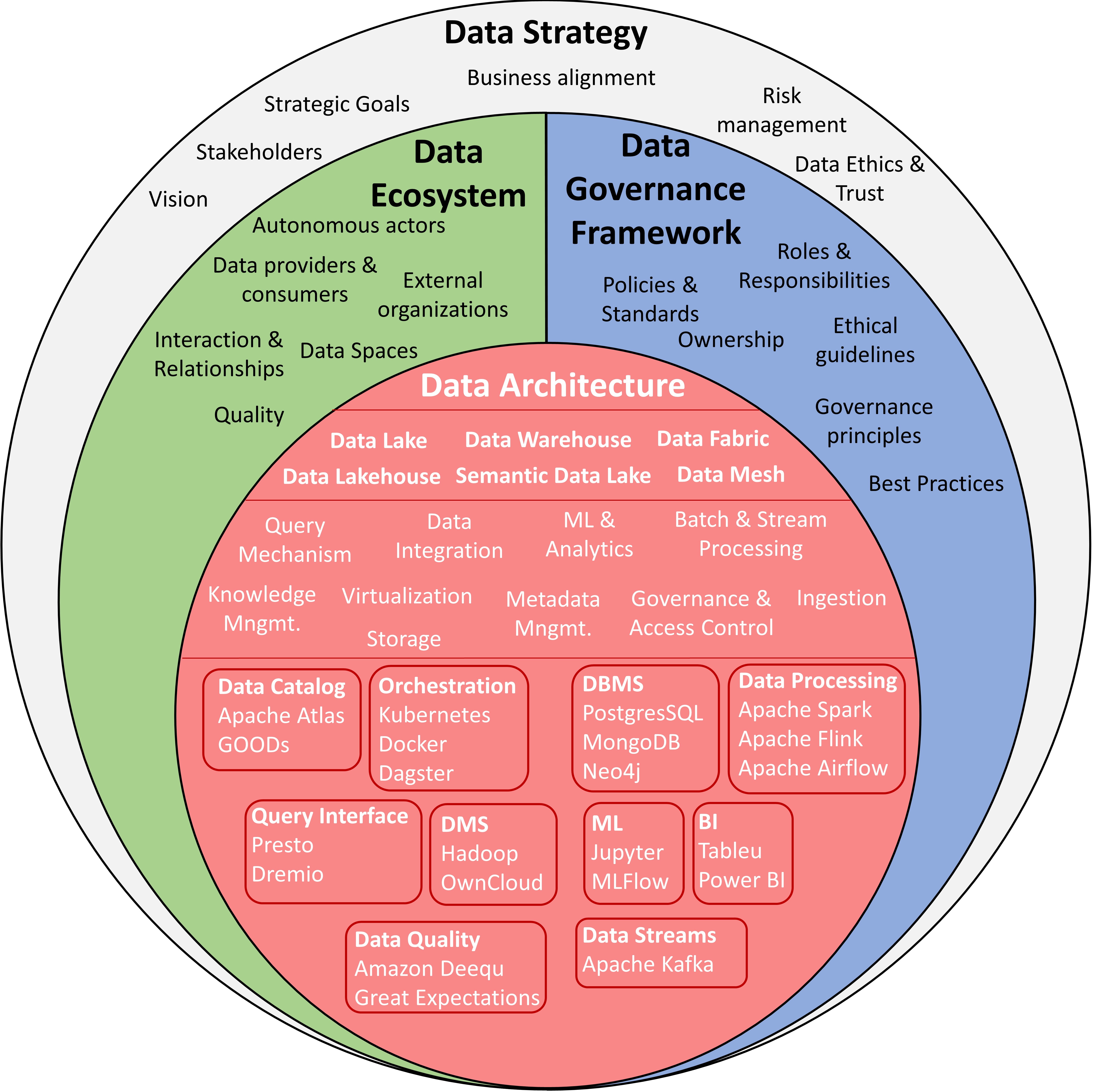}
		\caption{Overview of terminology, abstractions and components for data architectures}
		\label{fig:terminology}
\end{figure*}

Terms such as architecture, framework, ecosystem, platform and infrastructure are strongly related and often used interchangeably, causing confusion and misinterpretation \cite{giebler2021data}. 
For example, Geisler et al. \cite{geisler2021knowledge} introduce and define the term data ecosystem, but point to \textit{Ontario} \cite{Endris2019} and \textit{Constance} \cite{hai2016constance} as concrete implementations, which are titled as (semantic) data lake by the authors.
While DAMA-DMBOK \cite{international2017dama} distinguishes between data architecture, framework, and strategy, their descriptions occasionally overlap, reflecting the inherent ambiguity of terminology in this domain.
\Cref{fig:terminology} illustrates the introduced terminology in a hierarchy of data strategy, ecosystems, governance frameworks, architectures, and technical components. 
Each concentric ring moves from abstract, organizational-level concerns (such as strategy and governance) toward concrete, technical implementations, e.g., data architectures and individual components for data processing.
The inner layers in the core of the figure highlight architectural aspects, grouped first by the selected architectures in this article, then by central technical functionality, and finally with a selection of concrete technologies for individual components. 

The concepts data strategy, data ecosystems, data governance framework and data architecture describe different perspectives of data management. 
In the following, we aim to give precise definitions, while being aware that others may use the terms differently depending on the context: \\
\textbf{Data strategy} refers to a high-level plan for how an organization will collect, manage, and use data to achieve goals, which must be directly aligned with the overall business strategy \cite{international2017dama}. \\
\textbf{Data ecosystem}, as defined by \cite{10.1145/3209281.3209335}, refers to ``\textit{set of networks composed of autonomous actors, which consume, produce, or provide data or other related resources}''. 
It describes the information management landscape of an organization on a conceptual level, i.e., how information is used or produced in business processes as well as how it is managed in information systems.
It might include relationships to external organizations, e.g., for information exchange along the supply chain. \\
A \textbf{data governance framework}\phantomsection\label{data_governance_framework} defines best practices, standardized roles, processes, technologies, and guiding principles for managing data assets across their lifecycle, in alignment with the data strategy of the organization. 
It defines the governance principles of the data ecosystem of an organization. \\
\textbf{Data architecture} refers to the technical design of multiple data-related components and their interactions. It should be developed according to the guidelines and goals defined in the data strategy and governance framework. \\
\textbf{Data platform}, in this article, refers to a concrete implementation of a data architecture.

To clearly distinguish each term, we address potential points of confusion by discussing areas where their meanings may overlap.

Data ecosystems are usually composed of multiple data architectures, i.e., an organization might have data warehouses as well as data lakes. The abstraction at data ecosystem level should not only focus on technical aspects and tools, it should also include people and policies. 
In this sense, the term ecosystem emphasizes the socio-technical dimension, focusing on actors, governance structures, interoperability standards, and shared infrastructure.

In this sense, we illustrated data governance frameworks as complementary to data ecosystems. Data ecosystems rather focus on actors, components, and their relationships, whereas a data governance framework aims at policies, standards, responsibilities, and guidelines. 

The core of \Cref{fig:terminology} illustrates common functions and components of data architectures.
For example, we view database management systems, data management systems, and data catalogs as individual components of data architectures.
We distinguish between \emph{Database Management Systems} (DBMSs) and the broader category of \emph{Data Management Systems} (DMSs).
A DBMS is a software system that supports the definition, querying, and manipulation of structured data, typically offering declarative query languages (e.g., SQL), transaction management, and consistency guarantees \cite{date1977introduction}. DBMSs, but also DMSs, are central components of many data architectures and are usually tailored to a specific data model, which defines their query capabilities.

While traditional relational DBMSs remain dominant, recent years have seen the rise of specialized systems:
\begin{itemize}[leftmargin=10pt]
    \item NoSQL DBMSs, categorized into document-oriented (e.g., MongoDB), column-oriented (e.g., Cassandra, HBase), graph-oriented (e.g., Neo4j), and key–value stores \cite{fathy2019unified},
    \item Vector DBMSs (e.g., QDrant) for similarity search and AI workloads,
    \item Time-series DBMSs (e.g., InfluxDB) for stream and sensor data, and
    \item Search-oriented DBMSs (e.g., Elasticsearch) for full-text indexing and distributed search.
\end{itemize}

In contrast, a \emph{Data Management System} (DMS) provides basic mechanisms for storing, accessing, and organizing data, often in the form of files or objects, but lacks advanced DBMS features such as declarative query languages, fine-grained transaction support, or schema enforcement. DMSs prioritize scalability and availability over strict structure, and typically provide only basic access methods such as file download. Examples include Apache Hadoop and Ceph.
Another important component of a data architecture is a \emph{Metadata Management System} or \emph{Data Catalog}, which is a system used to manage and organize metadata.
Data catalogs provide capabilities for tracking data lineage, enabling data search and discovery, and ensuring that data is well-documented and accessible for various users. They play a crucial role in data governance and help organizations maintain compliance, ensure data quality, and enhance the overall usability of data.

\section{Data Architectures}\label{section:data_architectures}
With the terminology, we now introduce and analyze six prominent data architectures: data warehouse (DW), data lake (DL), semantic data lake (SDL), data lakehouse (DLH), data fabric (DF), and data mesh (DM). 
Each architecture is presented with a brief discussion of its historical motivation, followed by a definition, characteristic features, formal requirements, and an architecture diagram.

\begin{figure*}[t]
		\centering
		\fbox{\includegraphics[width=0.7\textwidth]{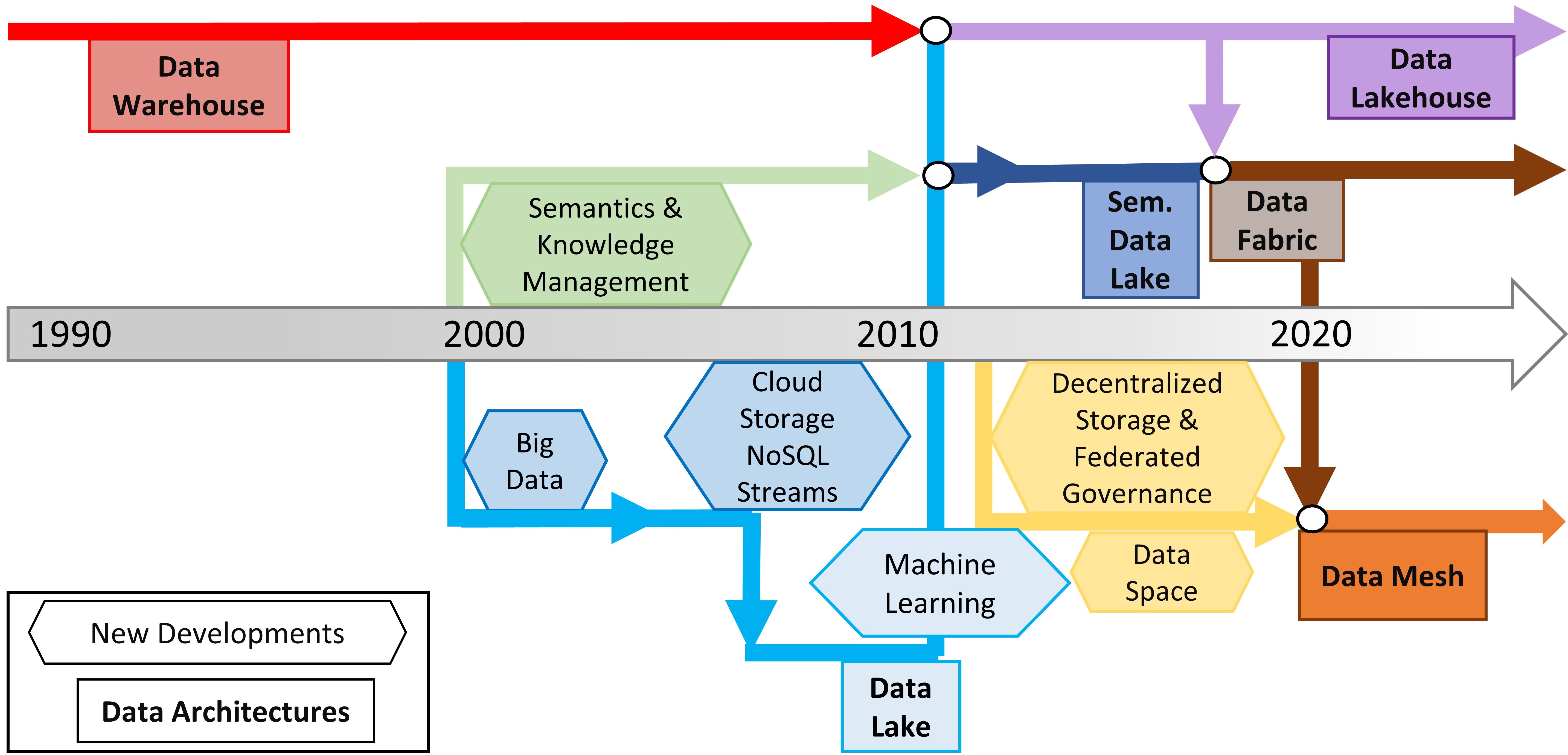}}
		\caption{Evolution of the various discussed data architectures}
		\label{fig:data_architecture_evolution}
\end{figure*}

\Cref{fig:data_architecture_evolution} illustrates the general historical development and conceptual relationships among these architectures. 
The evolution begins with data warehouses, which emerged in the 1990s to support analytical reporting and business intelligence through centralized, materialized, and integrated repositories \cite{inmon2005building}. 
An important precursor was the federated database system of the 1980s, which offered virtual integration of heterogeneous databases without physically centralizing the data \cite{10.1145/96602.96604}. 
The Semantic Web, introduced around 2000 \cite{DBLP:journals/internet/DeckerMHFKBEH00}, laid the groundwork for knowledge representation and semantic integration. 
Around the same time, the term \emph{big data} was introduced \cite{laney20013d}, emphasizing challenges in volume, velocity, and variety.

The 2000s and early 2010s saw a range of technological shifts: the rise of \emph{NoSQL} database systems to address schema flexibility and scalability \cite{moniruzzaman2013nosql}, the emergence of cloud storage platforms such as Amazon S3 \cite{armbrust2010view}, and the increasing relevance of data stream processing for real-time analytics \cite{golab2022data}. 
The machine learning (ML) boom, accelerated by advances in deep learning from 2012 onward \cite{krizhevsky2012imagenet,simonyan2014very,he2016deep}, further increased demand for architectures capable of handling heterogeneous, high-volume, and rapidly changing data \cite{putrama2024heterogeneous}. 
In response, data lakes emerged around 2010-2015 \cite{dixon2010pentaho}, offering scalable storage with schema-on-read and ELT paradigms. 
These were later extended into semantic data lakes, which incorporate ontologies and metadata for improved data understanding and integration \cite{hoseini2023semantic}.

As data became increasingly distributed across departments, organizations, and jurisdictions, new architectural paradigms emerged to address governance, sovereignty, and scalability. 
Data mesh introduces domain ownership and a ``data-as-a-product'' philosophy to promote scalability and accountability within organizations \cite{dehghani2022data}. 
Around 2019, data fabric was proposed \cite{zaidi2019data} as a metadata-driven integration layer that leverages AI for governance and enables self-service access across heterogeneous environments. 
More recently, the data lakehouse concept has gained traction \cite{armbrust2021lakehouse}, combining the reliability and ACID guarantees of data warehouses with the flexibility of data lakes to support unified batch and real-time analytics.

Overall, the figure highlights a clear evolution from centralized and structured systems toward decentralized, federated, and semantically enriched architectures. This trend reflects the increasing heterogeneity of data sources and the distributed nature of modern enterprises. 
Nevertheless, data warehouses, lakes, and lakehouses remain central in practice due to their maturity and strong theoretical foundations; especially in scenarios where decentralization is not a primary requirement and user groups are relatively homogeneous \cite{dehghani2022data}.

\subsection{Data Warehouse}
\label{section:data_warehouse}

In the 1990s, data warehouses \cite{inmon2005building} emerged as a standard solution for enterprise-wide analytical data management. They enabled organizations to run overnight ETL processes (Extract-Transform-Load) that extracted data from heterogeneous sources, transformed it to meet analytical requirements, and loaded it into a centralized, materialized repository. This architecture supported business intelligence by providing a consistent and integrated view of enterprise data.

Traditionally based on relational database technology, data warehouses are the most established form of analytical data architecture \cite{inmon2005building} and remain widely used in industry. They have evolved to support cloud-native implementations that offer greater scalability and lower operational costs. For instance, \emph{Apache Hive} \cite{huai2014major} introduced data warehouse functionality for big data by enabling SQL queries on top of \emph{Hadoop}, while \emph{Snowflake} \cite{DBLP:journals/dbsk/HentschelDFHO25} exemplifies a commercial, cloud-native data warehouse. Despite these innovations, data warehouses primarily support predefined analytical queries for reporting purposes and are less suited for exploratory or unstructured analytics such as text or web mining \cite{chaudhuri1997overview,bose2009advanced}.

\subsubsection{Characteristics}

Data warehouses are designed to manage structured data and often rely on multi-dimensional data models for complex analytical tasks. They typically ensure atomicity, consistency, isolation, and durability (ACID) properties and offer advanced capabilities such as time travel, data governance, and zero-copy cloning \cite{armbrust2021lakehouse}.

For the definition we rely on Inmon \cite{inmon2005building}:
\begin{definition}\label{definition-datawarehouse}[\textbf{Data Warehouse}]
	"\textit{A data warehouse is a subject oriented, integrated, non-volatile, and time variant"}
		data architecture \textit{"in support of management’s decisions.}" \cite{inmon2005building}
\end{definition}

\emph{Subject-oriented} refers to the purposeful organization of data to support business analysis and decision-making.
\emph{Integration} is the central aim: although data is collected from diverse systems, the warehouse presents a unified view to the user.
\emph{Non-volatility} indicates that data is not regularly updated by operational systems but remains stable to support long-term analysis.
\emph{Time-variance} reflects the inclusion of historical data, enabling trend analysis and retrospective insights.

In practice, the term ``data warehouse'' may also refer to centralized repositories that include little or no historical data. However, the core idea remains that it provides business users and decision-makers with consistent, reliable access to analytical data \cite{harby2024data}.

\begin{figure}[tb]
		\centering
		\includegraphics[width=0.7\columnwidth]{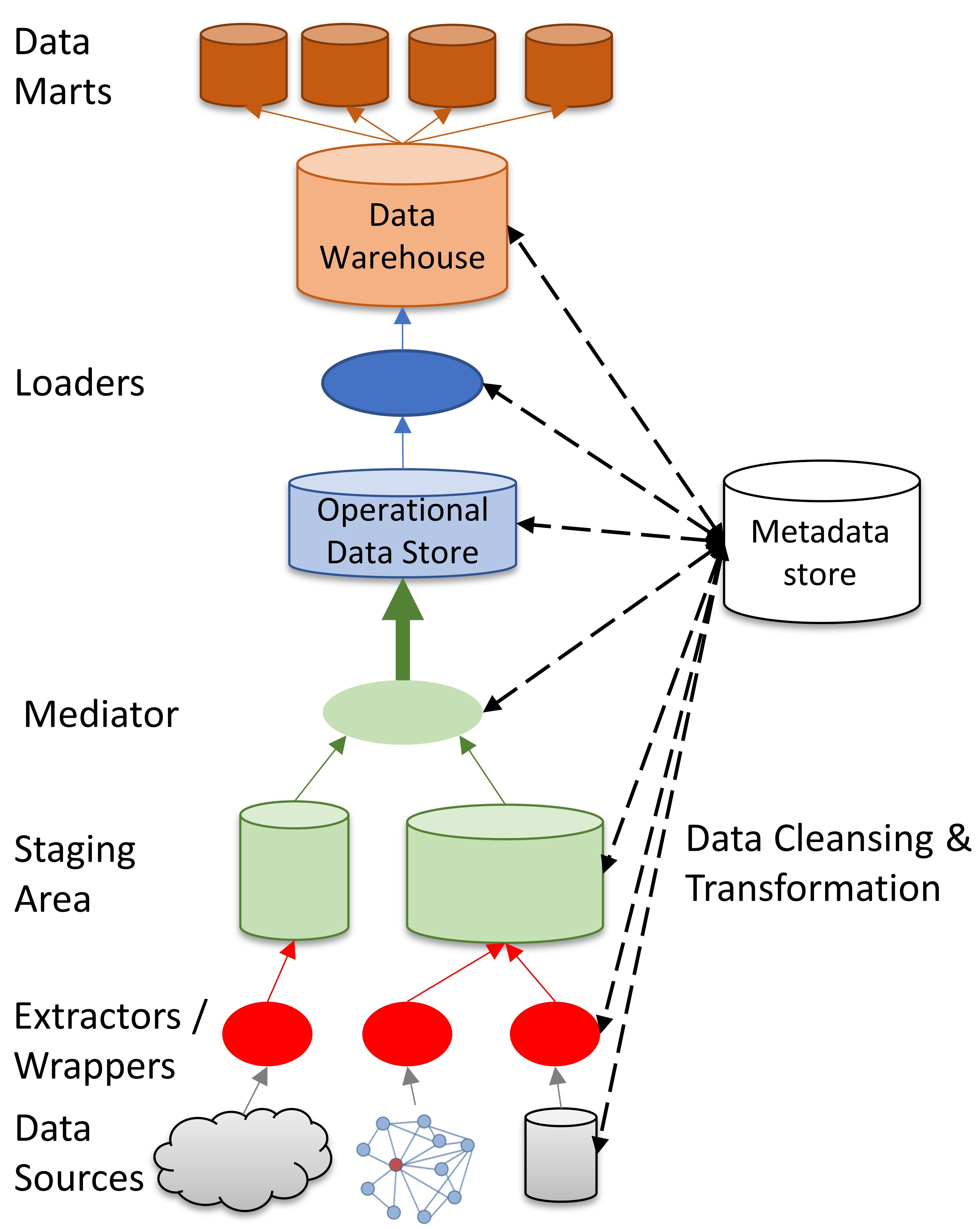}
		\caption{Data warehouse architecture based on \cite{JJQV99IS,DWQBuch}}
		\label{fig:data_warehouse_architecture}
\end{figure}

The architecture, illustrated in \Cref{fig:data_warehouse_architecture} \cite{JJQV99IS,kimball2013data,putrama2024heterogeneous}, revolves around ETL processes that consolidate data from various sources. While the core storage is generally relational, the staging area may involve other systems, including NoSQL stores \cite{davoudian2018survey}. To support performance and modular access, data marts are employed as thematic subsets of the warehouse, optimized for \emph{Online Analytical Processing} (OLAP) \cite{chaudhuri1997overview}.
Data sources are independent systems from which data is extracted. The staging area serves as a buffer for cleansing and transforming data before integration. A mediator integrates this harmonized data into a consistent, unified store. The operational data store holds historical data independently of specific use cases. The data warehouse acts as the central, structured repository optimized for analysis. Data marts are thematic subsets designed to support modular and secure data access. Finally, the metadata system manages schema and lineage information, enabling effective data governance and understanding.

\subsubsection{Requirements}

The following technical requirements capture the essential features of a data warehouse (\cite{inmon2005building,chaudhuri1997overview,DWQBuch}):
\begin{description}
	\item[DW1 ---]\phantomsection\label{DW1} \textbf{Schema-on-write}: data must conform to a predefined schema before it is written and stored. 
    Hence, an ETL process consolidating data from heterogeneous sources is required to transform the data into the desired target schema.

	\item[DW2 ---]\phantomsection\label{DW2} \textbf{Structured data model}: data is organized into a structured schema following a relational or multi-dimensional model.

	\item[DW3 ---]\phantomsection\label{DW3} \textbf{Consistency guarantees}: provide means for checking and enforcing the consistency of structure and content across data collections.

	\item[DW4 ---]\phantomsection\label{DW4} \textbf{Optimize for read-heavy workloads}: optimize query performance, making it ideal for analytical and reporting tasks.

	\item[DW5 ---]\phantomsection\label{DW5} \textbf{Time-variant data storage}: include historical data, supporting trends and time-based analysis by retaining snapshots of data over time.

    \item[DW6 ---]\phantomsection\label{DW6} \textbf{Query language}: offer a declarative, structured query language like SQL.

    \item[DW7 ---]\phantomsection\label{DW7} \textbf{OLAP}: enable Online Analytical Processing (OLAP) by supporting multidimensional queries for business intelligence.
\end{description}

\subsection{Data Lake}\label{section:data_lake}
Since their emergence, numerous proposals have addressed data lake architectures, in particular, challenges, concepts, components, and implementations \cite{Quix2019,sawadogo2021data,scholly2021coining,hai2016constance,zhao2021data,farid2016clams:short,10.1145/3340531.3417426,Dibowski2020UsingST,DBLP:journals/dke/EichlerGGSM21}. 
In addition, major IT companies have introduced commercial tools for data lakes \cite{nambiar2022overview,ramakrishnan2017azure, DBLP:journals/debu/HalevyKNOPRW16}.
In the following, we focus on widely recognized concepts and representative architectural patterns discussed in current academic literature or industrial implementations \cite{azzabi2024data,hai2023data,10.14778/3352063.3352116}.

In 2010, James Dixon \cite{dixon2010pentaho} introduced the term data lake as a solution for managing raw data from a single source while accommodating diverse user needs, because ``\textit{data is of a scale or daily volume such that it won’t fit technically and/or economically into a relational DBMS''} \cite{dixon2010pentaho}.
This concept stood in stark contrast to the so-far dominant data warehouse systems.
By preserving data in its raw format, data lakes bypass or delay the costly preprocessing steps. 
Dixon motivates the choice for the word lake with an analogy:
``\textit{If you think of a datamart as a store of bottled water -- cleansed and packaged and structured for easy consumption -- the data lake is a large body of water in a more natural state. 
}'' \cite{dixon2010pentaho}.

In 2014, Gartner \cite{gartner2014data} raised several criticisms of data lakes, warning that the indiscriminate ingestion of disparate data could lead to a so-called data swamp, making the data unusable without proper metadata management and governance \cite{Quix2019}.
Further, concerns included insufficient attention to data security and privacy oversight. 
In response to these critiques, Dixon revisited the data lake concept \cite{dixon2014revisited}.

With Hadoop at the core, a key requirement was that a data lake should offer low-cost, easily accessible storage that supports a \textbf{schema-on-read} approach, meaning the metadata can be populated over time and transformations to apply the schema are only performed once required. 
This approach allows a data lake to automatically extract metadata from raw data and identify patterns, with users contributing additional descriptive information, such as domain-specific knowledge and attribute linkages. 
Hence, the idea is that the interaction between users and the data lake continually enhances (meta)data quality and value.

\begin{figure*}[h]
    \centering
    \begin{minipage}{0.45\textwidth}

        \includegraphics[width=\linewidth]{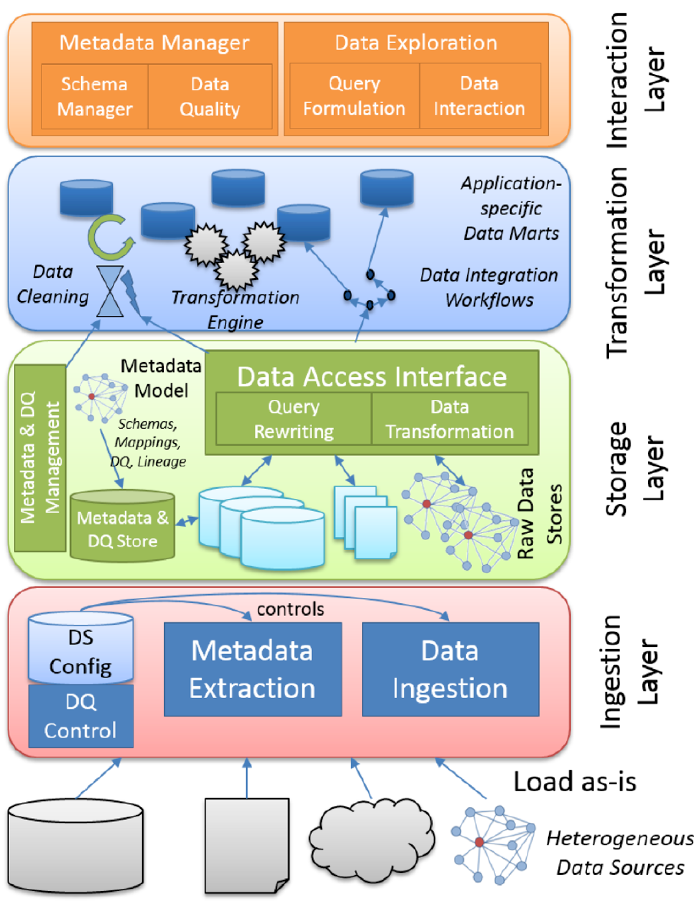}
        \caption{Four-layered data lake architecture \cite{DBLP:conf/birthday/JarkeQ17}}
        \label{fig:datalake}
    \end{minipage}
    \hfill
    \begin{minipage}{0.45\textwidth}
        \includegraphics[width=\linewidth]{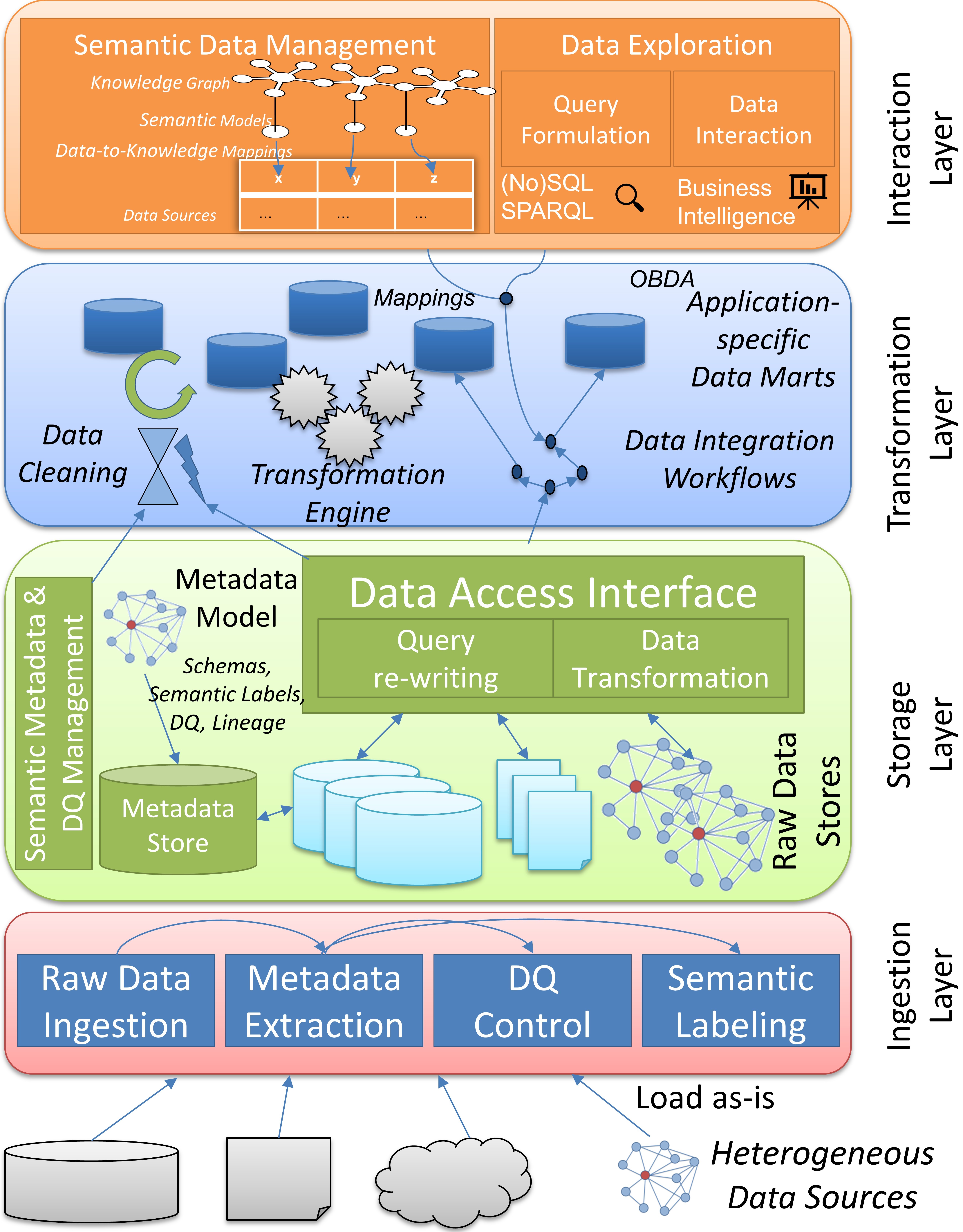}
        \caption{Semantic data lake architecture \cite{hoseini2023semantic}}
        \label{fig:semantic_data-lake-arch}
    \end{minipage}
\end{figure*}


\subsubsection{Characteristics} \phantomsection\label{ETL}

The classical ETL approach was to limited to handle the requirements at the time, because transforming data to fixed schemas and rigid data models leads to limited flexibility in data management. Additionally, some information might be lost during the process, as it might not fit into the target schema and scaling often required a substantial financial investment, both in terms of initial hard- \& software costs and ongoing maintenance \cite{azzabi2024data}.

In contrast, data lakes load \textbf{raw data} from multiple sources into a unified management system with a heterogeneous storage layer. 
They provide flexible data access, automated and efficient data preparation workflows, and a series of data pipelines for a wide range of analytical applications of which end-to-end machine learning workflows become increasingly important \cite{hai2023data,hoseini2024enhancing}.
Data is transformed only if required for a specific application or query (schema-on-read). 
Thus, data lakes apply an Extract-Load-Transform (\textbf{ELT}) approach. 
As data heterogeneity and the number of sources increase, data lakes follow a data-first,
schema-on-read approach, reflecting the shift from frequent, schema-driven access toward more
flexible, on-demand usage \cite{bacco2024data}.

Query processing needs to deal with heterogeneous data sources in an on-the-fly manner \cite{hai2023data}, as data is not transformed into a uniform format before it is loaded. 

Further, data lakes are defined by their ability to scale to massive volumes, all while offering \textbf{low-cost storage} solutions typically based on Hadoop, enabling organizations to store petabytes of data \cite{azzabi2024data}. 

\textit{Apache Spark} is a system frequently used for the implementation because it can process queries over heterogeneous data sources.
For example \textit{Constance} \cite{hai2016constance} implements a polystore by storing diverse raw data according to its original format: relational (e.g., \textit{MySQL}), document-based (e.g., \textit{MongoDB}), and graph databases (e.g., \textit{Neo4j}). 

A key feature is the use of metadata to manage, organize, and analyze the stored data, because otherwise the data is hardly usable as their structure and semantics are not known. 
Hence, \textbf{metadata extraction} or \textbf{metadata enrichment}, the process of automatically discovering metadata information of a dataset, is essential for accessing datasets in a later phase. 
This includes all types of metadata, the more the better: structural, but also descriptive information as well as relationships to other datasets.
Metadata models answer the question of which information to collect and how to structure and organize it and there numerous proposals in recent years \cite{hoseini2024enhancing,scholly2021coining,Quix2016,sawadogo2019metadata,eichler2020handle, ouellette2021ronin}.

Additionally, data lakes aim to enable data exploration and discovery, allowing analysts to navigate and examine heterogeneous data sources without needing predefined schemas or structures and this is a complex challenge and subject to ongoing research \cite{paton2023dataset,hoseini2023sedar,DBLP:conf/semweb/MamiGSJA019a,Endris2019,Bagozi2019}. 
Rencently, the increasing need for supporting machine learning processes inside a data lake system is emphasized \cite{hai2023data,10.14778/3352063.3352116}.

We adopt our definition for data lakes from \cite{hoseini2023semantic} and added further technical requirements:
\begin{definition}\label{definition-datalake}[\textbf{Data lake}]
	A data lake is a data architecture that manages heterogeneous data sets in their original structures. It should provide functions to extract data and metadata from heterogeneous sources, manage the data efficiently in heterogeneous storage systems, and query and transform the data in a scalable way. Furthermore, a uniform query processing mechanism is necessary, enabling users to query data seamlessly across diverse formats and sources without requiring extensive pre-processing or schema transformations  \cite{Quix2019,hai2023data}.
\end{definition}

\subsubsection{Requirements}

\begin{description}
	\item[DL1 ---]\phantomsection\label{DL1} \textbf{Schema-on-read}: allow raw data of any format to be stored. Schema definitions are only required on access to the data.
	
	\item[DL2 ---]\phantomsection\label{DL2} \textbf{Data ingestion at scale}: support diverse data ingestion methods, such as batch processing, streaming, and real-time data flow for arbitrary heterogeneous data sources.
	
	\item[DL3 ---]\phantomsection\label{DL3} \textbf{Scalability and cost-effective storage}: handle vast amounts of data with elasticity to expand or contract storage based on needs for cost efficiency.
	
	\item[DL4 ---]\phantomsection\label{DL4} \textbf{Data governance and metadata management}: provide a robust data catalog with automated metadata extraction, searchable indexes, tags, and descriptions for datasets ensuring data quality, lineage for data discovery to locate, understand, and utilize data independently of analysis.
	
	
	\item[DL5 ---]\phantomsection\label{DL5} \textbf{Unified query mechanism}: enable querying of heterogeneous data sources using a unified query interface allowing users to perform federated queries across structured and unstructured data with on-demand data transformations.
\end{description}

Azzabi et al. \cite{azzabi2024data} outline how the initial \textit{mono-zone architecture} had notable limitations and hence data pond and zone-based architectures were introduced.
Zone-based architectures 
include transient landing, raw, trusted, and refined zones, along with a sandbox zone for exploratory analysis. 
Each zone has a specific function based on its refinement level, from raw data ingestion and preliminary processing to serving compliant, quality-checked data to end-users \cite{sharma2018architectingdatalakes,patel2017datalakegovernance,ravat2019data,madsen2015enterprisedatalake,zikopoulos2015bigdata,gorelik2016enterprise}.  
A \textit{four-layered data lake architecture} was proposed in \cite{DBLP:conf/birthday/JarkeQ17} and is shown in \cref{fig:datalake}. It resembles some ideas of zone-based architectures, such as a raw data `zone' at the storage layer, and refined data in data marts at the transformation layer. In addition, it emphasizes the role of metadata and quality management on each of the layers.

The \textit{Ingestion Layer} loads data ``as-is'' into the data lake while maintaining its original format and structure thereby facilitates the acquisition of raw, heterogeneous data from various data sources, including databases, files, and streaming platforms. It is responsible for metadata extraction to provide insights into the structure, quality, and other properties of the incoming data.

The \textit{Storage Layer} provides a centralized repository for storing raw data and metadata retaining the original format of ingested data.
It manages metadata and quality information, enabling efficient querying and processing and ensures that raw data remains immutable, supporting traceability and reproducibility.

The \textit{Transformation Layer} is responsible for the integration and preparation of data transforming raw data into refined datasets to meet analytical requirements, it facilitates query rewriting and data transformation tasks, such as data cleaning, enabling seamless data retrieval and manipulation.

The \textit{Interaction Layer} focuses on enhancing the accessibility and usability for end-users acting as the user-facing interface. 
It comes with some form of metadata manager that handles schema mapping, data quality assessment, and metadata management as well as data exploration to enables users to explore datasets, query the data lake, and interact with application-specific workflows.

\subsection{Semantic Data Lake}\label{section:semanitc_data_lake}
Various strategies for the use of semantic data management have been introduced \cite{Dibowski2020UsingST,Paulus2021,hoseini2023semantic,putrama2024heterogeneous}, which underscore the necessity of integrating big data with semantic web technologies into \textit{semantic data lakes}.
A semantic data lake facilitates the derivation semantic models from raw datasets, thus providing a more conceptual data description that includes concepts and their interrelations, and linking data sets to their relevant concepts \cite{Dibowski2020a}. Recently, with the growing demand for knowledge management, knowledge graphs have also been integrated into data lake architectures \cite{dibowski2021using,DBLP:journals/pvldb/BeheshtiBNT18}.

At the company \textit{Bosch}, the potential of semantic technologies for data management has already been exploited at scale in production \cite{DBLP:conf/semweb/KalayciGLXMKC20}.  
Dibowski et al.~\cite{dibowski2021using, Dibowski2020a} describe a metadata model represented as an ontology based on W3C recommendations for the data catalog management and provenance control of a semantic data lake.
Ingested data assets are \textbf{enriched with semantics} by aligning, annotating, and enriching the input data with semantics.
The development of knowledge graphs required an immense initial effort by the company but promised a long-term solution for a series of sophisticated challenges. 
In addition, they utilize a new concept called ontology-driven, self-adaptive frontends, in which user interfaces for data catalogs can dynamically render information from a knowledge graph. A similar concept was presented in \cite{Calvanese2021}. 
This is useful because changes in the underlying ontology do not require any changes at the frontend.
The \textit{ESKAPE} system from 2017 \cite{Pomp2017} can be viewed as a very first prototype in this direction and the term was coined by Pomp et al. \cite{pomp2018applying} and Mami et al. \cite{mami2018teach} for the first time in 2018.

\subsubsection{Characteristics}
We presented a thorough review of the field in \cite{hoseini2023semantic} from which we derive the following definition and requirements:
\begin{definition}\label{definition-semanticdatalake}[\textbf{Semantic data lake}]
	Semantic data lakes are a specific form of traditional data lakes that extend the capabilities through a \textit{semantic layer} that enriches and connects the stored data \textit{semantically}. 
	The \textit{semantic layer} equips data sets in the lake with connections from the data set's metadata to conceptual/logical models, which encapsulate knowledge potentially external to the content of the data, such as domain knowledge. 
\end{definition}

\subsubsection{Requirements}
The following, together with the requirements for a traditional data lake (see \cref{section:data_lake}), confine the semantic data lake.
\begin{description}
	\item[SDL1 ---]\phantomsection\label{SDL1} \textbf{Creation of semantic labels}: enable the mapping of data attributes and other schema elements to elements of a KG (might be internal or external, domain-specific).
	\item[SDL2 ---]\phantomsection\label{SDL2} \textbf{Semantic relationships}: represent any form of hierarchical, generic, or pre-defined semantic relationships (semantic connections between data sets, e.g., for provenance or governance).
	\item[SDL3 ---]\phantomsection\label{SDL3} \textbf{Metamodel as ontology}: the metamodel of the lake itself is modeled as ontology using semantic web technologies.
    \item[SDL4 ---]\phantomsection\label{SDL4} \textbf{Semantics-based data access}: the ability to query and access data using semantic-based query languages like SPARQL, i.e., abstracting from physical storage and allowing to interact with the data conceptually using techniques like ontology-based data access \cite{DBLP:conf/ijcai/XiaoCKLPRZ18}.
\end{description}
In \cite{hoseini2023semantic}, we introduce a semantic data lake architecture (see \cref{fig:semantic_data-lake-arch}) based on the four-layered architecture. 
Dibowski et al. \cite{dibowski2021using} describe a semantic data lake as a ``\textit{specific form of data lakes in which a semantic layer on top enriches and connects the data semantically}''. 
This definition suggests to extend the four-layered architecture by a fifth layer. In our view, proper semantic data management is only possible if it is an integral part of the data architecture.
Therefore, incorporating semantics demands adjustment across the entire functionality, starting possibly already at the ingestion, where users (or AI-based technologies) annotate a dataset with semantic labels, before it is fully available inside the lake for accessing. 
Thus, each layer of the architecture has been extended with enhanced metadata functions with semantic capabilities. 
For instance, semantic labeling in the ingestion layer assigns semantic labels to metadata elements. 
Data quality management benefits from an expanded semantic metadata repository managed within the storage layer, which supports functions such as ontology-based data access (OBDA) \cite{hoseini2023semantic}, aiding in data usage and interpretation. 
Consequently, the interaction layer incorporates additional features, such as exploration of the knowledge graph and the semantic model, semantic query formulation using languages like SPARQL, and tools for refining semantic mappings and models.

\subsection{Data Lakehouse}\label{section:data_lakehouse}

Although data warehouses and data lakes serve different analytical purposes, organizations frequently require features of both worlds simultaneously \cite{groger2021there}. Especially the necessity of generating business reports and crafting ML models in a single platform gave rise to the \textit{data lakehouse}.
Armbrust et al. \cite{armbrust2021lakehouse} recognize that many organizations host an expensive
two-tier architecture which is highly complex for users. 
Data is first ingested into lakes, and then again fed via an ETL process into data warehouses, creating complexity, delays, and new failure modes. 
Consistency is one of the challenges, i.e., keeping the data lake and warehouse in a consistent state is difficult and costly.
Data is stale, new data is loaded to the data warehouse only with a delay due to the ETL process. 
Finally, both architectures provide limited support for advanced analytics. 

The typical data warehouse supports well-designed business intelligence (BI) queries, which are known in advance. 
The multidimensional data model of data warehouse is optimized to support these queries. 
However, ad-hoc queries to extract large of amounts of data for a new purpose, e.g., training data for ML models, are less efficient in a data warehouse context.
Data lakes, on the other hand, lack the rich and matured features from data warehouses, such as ACID transactions, data versioning and indexing \cite{armbrust2021lakehouse}.
In response, Databricks, the company founded by the creators of \textit{Apache Spark}, introduced the lakehouse architecture to combine the best features of both systems. 
Building on their expertise with Spark's scalable data processing capabilities, they introduced \textit{Delta Lake} in 2019 \cite{armbrust2020delta} as an open-source storage layer that brings ACID transactions, schema enforcement, and data versioning to data lakes. 

\subsubsection{Characteristics}
Since key characteristics of data warehouses and data lakes often conflict (particularly in terms of data access, data independence, and storage formats) combining them typically requires compromises on both sides. 
For instance, a data warehouse that supports direct access to storage, as in data lakes, must sacrifice data independence and adopt open file formats. 
Conversely, a data lake can introduce warehouse-like management features only by restricting users' flexibility in storing and accessing data, enforcing structured protocols for read and write operations \cite{Schneider2023}.
The data lakehouse architecture merges the flexibility of data lakes, which support the large-scale storage of various data formats, with the transactional integrity of data warehouses by incorporating layered components. 

Characteristic for a data lakehouse is a \textbf{transactional layer}, which ensures data integrity through ACID properties (Atomicity, Consistency, Isolation, Durability). 
This layer organizes the data and manages metadata, enforcing schema, supporting data versioning, and ensuring data lineage, which enhances overall data quality.
Here, technologies like Delta Lake, Apache Hudi, and Iceberg used together with Spark on top of scalable and cloud-native storages like HDFS form the core \cite{schneider2024lakehouse}.
Thus, the lakehouse has a \textbf{relational format} as its final structure, on top of heterogeneous sources, as most analytical and visualization tools are optimized for relational data, enabling faster analysis. 
Schneider et al. \cite{schneider2024lakehouse} conduct a thorough review on the data lakehouse and find four common definitions in current literature and argue that each one of them has certain shortcomings.
Hence, they propose the following and derive technical requirements from it. The following are direct quotes from \cite{schneider2024lakehouse}:
\begin{definition}\label{definition-datalakehouse}[\textbf{Data Lakehouse}]
	\textit{``A data lakehouse is a''} data architecture \textit{``that leverages the same storage type and data format for reporting and OLAP, as for data mining and machine learning, as well as streaming workloads''}.
\end{definition}

\subsubsection{Requirements}

\begin{description}
	\item[DLH1 ---]\phantomsection\label{DLH1} \textit{\textbf{"Same storage type and data format}: all (meta)data is stored on a single type of storage (no polystore), using a uniform open data format. Hence, a data lakehouse must have an inherently low heterogeneity in terms of the utilized technologies and data formats.}

	\item[DLH2 ---]\phantomsection\label{DLH2} \textit{\textbf{Relational data collections}: this strong requirement comes from the need to support reporting and OLAP workloads, which require a higher degree of structure than would be provided by other types of abstractions, such as graphs or documents.}

	\item[DLH3 ---]\phantomsection\label{DLH3} \textit{\textbf{Query language}: offer a declarative, structured data query language for the data collections.}

	\item[DLH4 ---]\phantomsection\label{DLH4} \textit{\textbf{ACID guarantees}: provide means for checking and enforcing the consistency of structure and content across data collections, and ensure isolation and atomicity for all operations that modify or access the data concurrently.}

	\item[DLH5 ---]\phantomsection\label{DLH5} \textit{\textbf{Direct read access}: provide unmediated, direct, and random read access to all (meta)data stored on the underlying storage system without needing to export and transform it first, and leverage open, standardized file formats.}
	\item[DLH6 ---]\phantomsection\label{DLH6} \textit{\textbf{Unified batch and stream processing}: support processing in large-scale batches as well as at a high rate and in real-time."}
\end{description}
\begin{figure}[t]
		\centering
		\includegraphics[width=0.9\columnwidth]{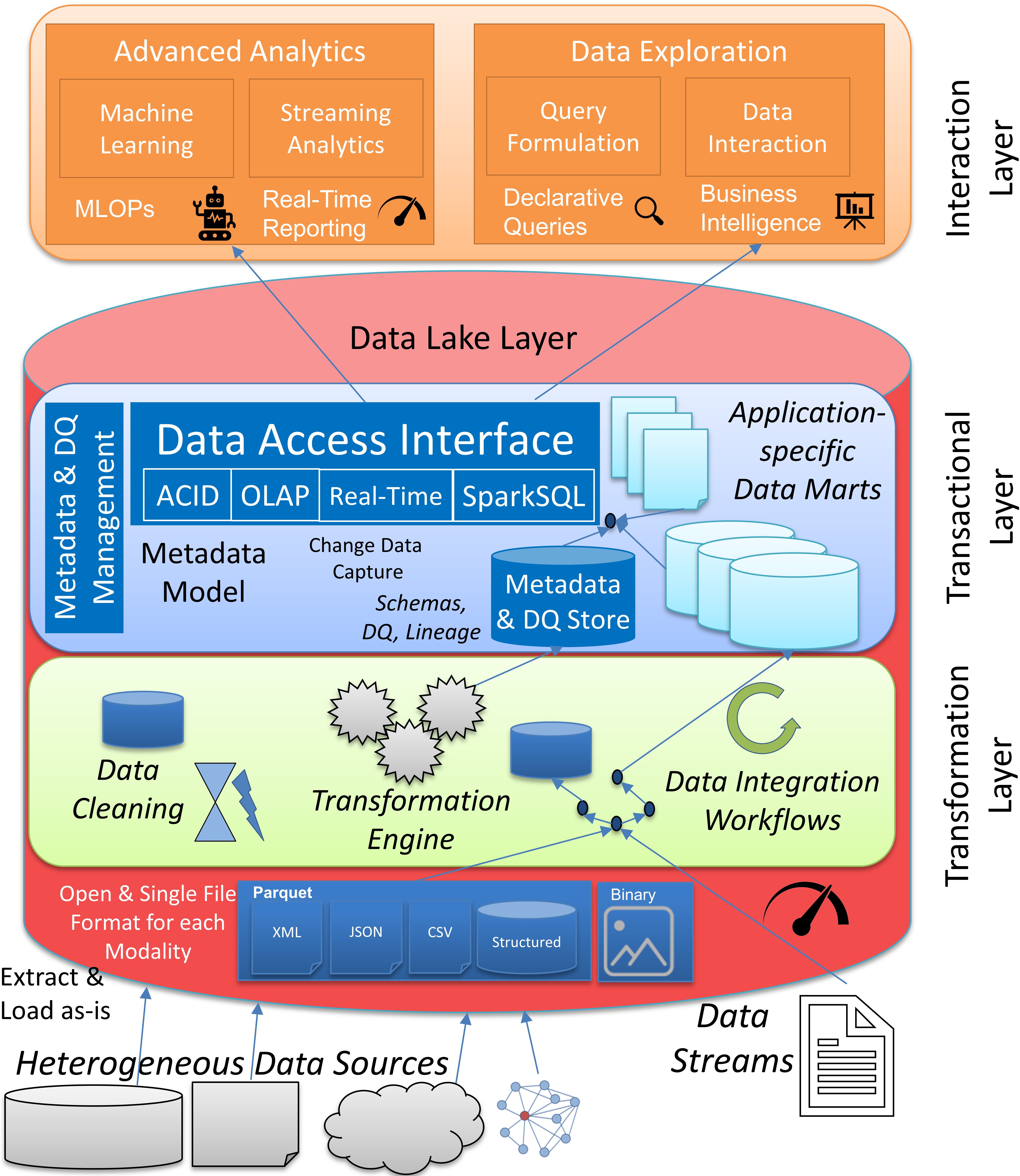}
		\caption{Data Lakehouse architecture inspired by \cite{DBLP:conf/birthday/JarkeQ17,armbrust2021lakehouse}}
		\label{fig:datalakehouse}
\end{figure} 
We derive a data lakehouse architecture (see \cref{fig:datalakehouse}) from the four-layered data lake architecture and the findings of Harby et al. \cite{10020719} and Azzabi et al. \cite{azzabi2024data}. 
In the figure, the ingestion layer is replaced by a composite \textit{Data Lake Layer}. 
This layer is composed of an ingestion layer, to load data and extract metadata, as well as a storage layer. 
What differs is that the storage is composed of single scalable storage technology, for example \textit{HDFS} or \textit{Amazon S3}.
This foundational layer serves as the primary storage for raw, structured, semi-structured, and unstructured data. 
It enables scalable, cost-effective data storage and accommodates diverse data sources, serving as a central repository for all ingested data.
Further, the layer utilizes a single open data format (such as Parquet) for each modality.

The \textit{Transformation Layer} is responsible for the ETL pipelines and thus dedicated to extracting data from various sources, transforming it into structured formats, and loading it for analysis. 
It standardizes and cleanses data to make it ready for analytics, enabling compatibility and consistency across data sources and preparing it for data access.
\textit{Transactional Layer} supports data operations and guarantees ACID compliance, enabling reliable and transactional processing, often needed for complex analytical queries (OLAP) supporting accurate and real-time analytical processing.
The transformation as well the transactional layer are embedded into the data lake layer to emphasize the vision that the lakehouse is a uniform system, and does not resemble a two-tier architecture composed of an individual data lake and data warehouse and thus no data replication or pipelines for transferring between the platforms are necessary.
Finally, the \textit{Interaction Layer} serves as the user interface ensuring that insights can be easily generated, accessed and utilized by business users and data scientists. 

\subsection{Data Fabric}\label{section:data_fabric}
The concept of data fabrics was coined by NetApp in a white paper from 2016 as a ``vision for data management […] that seamlessly connects different clouds, whether they are private, public, or hybrid environments'' \cite{hechler2023data}. Back then, the term meant a combination of different data storage and integration techniques, however, without details about a concrete architecture.

In 2019, Gartner defines a data fabric as a design concept for attaining reusable and augmented data integration services, data pipelines and semantics for flexible and integrated data delivery \cite{debellis2023interoperability}.
Hence, a data fabric is a technical architecture that brings together heterogeneous data that spans across multiple data sources; it allows organizations to monitor and manage data regardless of the location, considering appropriate data governance and data cataloging \cite{bode2024towards,li2022distributed,abu2023structural}.
Then in 2021, Gartner \cite{gartner_fabric} redefines a data fabric as a ``design concept that serves as an integrated layer (fabric) of data and connecting processes''. 
In their data fabric architecture, Gartner emphasizes key characteristics, such as the use of AI for metadata management and data cataloging as well as knowledge graphs and ontologies to connect and enrich data sources with business related concepts. 
They introduce the notion of \textbf{``passive vs active metadata''}, i.e., data models, schema definitions, glossary, various logs are considered as passive metadata, whereas the continuous analysis of data usage with are described as active metadata describing actual experiences.
The conversion of passive metadata into active metadata is interpreted as automating the collection and management of metadata \cite{debellis2023interoperability}.
By now, major vendors in data science have embraced the term; both vendors of semantic technology (e.g., Franz Inc., Ontotext, Pool Party, Stardog, and Top Quadrant) as well as more traditional vendors such as Microsoft, Oracle and Tibco \cite{debellis2023interoperability}.

\subsubsection{Characteristics}
At present, the particular mechanisms, standards, and technologies for the implementation of a data fabric architecture still have at be clarified as there is not a reference architeture or system \cite{hechler2023data,9671862,abu2023structural}.
For example, some authors advocate for the presence of \textbf{data virtualization}, i.e., a software architecture that allows applications and end users to access data from multiple sources as if it were stored in a single location, as a critical component of a data fabric \cite{kuftinova2022data,akermi2023data,li2022distributed}. In fact, the term ``fabric'' metaphorically represents the interwoven threads/fibers of a textile, signifying how different data sources, technologies, and workflows can be interconnected into a cohesive whole. 
As noted by \textit{CData Software} \cite{cdata2024}:  ``\textit{If we take the metaphor to its conclusion, data virtualization can weave the threads (data) to make the finished cloth (data fabric).}''  
Others see data virtualization as optional \cite{serra2024deciphering} or do not mention it at all \cite{macias2024data}. 
In our requirements definition below, we consider data virtualization as a key requirement for data fabrics. Data fabrics should be used as an additional integration layer on top of other existing data platforms.

Similarly, various proposals include the use of knowledge graphs, ontologies and semantic web technologies explicitly \cite{debellis2023interoperability} and others do not mention them at all \cite{rieyan2024advanced}. 
DeBellis et al. \cite{debellis2023interoperability} provide strong arguments in favor of using semantic web technologies because they ensure reuse and interoperability of data. 

Modeling metadata for a data fabric requires a complex, highly interconnected model that typically spans multiple organizational, geographic, and business boundaries. Obviously, relational data models are too rigid to easily store this type of data. 
Knowledge graphs are more suitable to represent such information because they are more flexible, heterogeneous, formal as well as fast and scalable \cite{debellis2023interoperability}. 
Martinez-Casanueva et al. \cite{martinez2024candil} present \textit{CANDIL}, a data architecture to convert heterogeneous data into an RDF-based representation using mappings. 
They call their platform a data fabric; however, besides the use of semantic data management techniques, we do not recognize other technical features in their approach. Therefore, we consider it as an ontology-based data integration pipeline.

A dominant theme in the data fabric paradigm is the idea of the \textbf{polystore}, i.e., combining heterogeneous storage databases, potentially hosted in the cloud \cite{9671862,strengholt2023data,macias2024data,alvord2022big,moon2021study}. 
Similarly, data storage is not meant to be centralized to a single location for access, but decentralized and distributed \cite{bode2024towards,serra2024deciphering}. 
Data exploration and access is mediated by the fabric through centralizing metadata only \cite{rieyan2024advanced,li2022distributed}.  

To facilitate this type of accessing, discovering, and understanding of data, but also for automating data integration and governance activities as well as privacy, and compliance-related topics, the data fabric puts metadata as the most critical element \cite{bode2024towards}.
Hechler et al. \cite{hechler2023data} differentiate the data fabrics paradigm from other data architectures by the predominant \textbf{use of AI to automate the metadata creation and management}, which is inline with the idea of active metadata.

According to Hechler et al. \cite{hechler2023data}, ensuring trustworthy AI by broadening the scope of traditional data governance toward unified AI governance is another requirement for data fabrics.
This is a relatively new area that goes beyond the inclusion of AI artefacts into the governance realm, because it requires to detect bias, drift, and decreasing accuracy and precision and needs to propose corrective actions to realign AI artefacts to business goals.

Furthermore, so-called self-service capabilities play a critical role that allow users to independently perform tasks that would traditionally require IT experts. 
Hechler et al. \cite{hechler2023data} provide concrete examples: (1) access various assets through the knowledge catalog based on roles, without needing additional credentials, (2) metadata includes details about how to access assets (e.g., (No)SQL, REST APIs, etc.), (3) a holistic view of AI/ML assets (4) metadata generation through AI to automate tasks like resource allocation and semantic annotations, (5) semantic search uncovers correlations and recommends relevant assets for business purposes, (7) data curation tasks such as data quality assessment and refinement are manageable by users without expert intervention.
These examples require the creation of standardized data pipelines based on comprehensive metadata management for the automated and systematic transformation, cleansing, and integration of data \cite{blohm2024data,moon2021study,macias2024data}.

An increasingly important pipeline is the support of the life cycle of AI processes \cite{hechler2023data}, i.e., the management of the in- \& outputs and the creation of ML models.
However, it is important to note that, although researchers have recognized the potential \cite{abu2023structural}, more traditional BI and data science activities, such as reporting, modeling and analysis tools are not considered the core scope of a data fabric, but rather fall under the responsibility of the data consumers \cite{9671862}.

For a definition we follow Blohm et al. \cite{blohm2024data}:
\begin{definition}\label{definition-datafabric}[\textbf{Data Fabric}]
	A Data fabric is a data architecture ``\textit{for attaining reusable data integration services, data pipelines and semantics for flexible and integrated data delivery \cite{zaidi2019data}. It builds on the analysis, creation, and usage of metadata  \cite{9671862,li2022distributed}, which are modeled as a knowledge graph.}'' \cite{blohm2024data}
\end{definition}

\subsubsection{Requirements}
Li et al. \cite{li2022distributed} identify key enabling technologies and Serra et al. \cite{serra2024deciphering} provide key features. Based on all of the previous findings we provide technical requirements for data fabrics: 

\begin{description}
	\item[DF1 ---]\phantomsection\label{DF1}  \textbf{Polystore and Decentralized Storage}: ensures data can reside across data lakes, data warehouses, or other systems, reducing redundancy and enhancing scalability.


	\item[DF2 ---]\phantomsection\label{DF2}  \textbf{Data Virtualization}: abstracts data complexity allowing seamless access to distributed sources via real-time querying and integration without requiring a central data repository.

	\item[DF3 ---]\phantomsection\label{DF3}  \textbf{AI-Driven Metadata Management}: metadata management must be (semi-)automated using AI to generate, enrich, and maintain metadata dynamically. This supports efficient data discovery, lineage tracking, and governance, while reducing manual intervention.

	\item[DF4 ---]\phantomsection\label{DF4}  \textbf{Self-Service Capabilities}: standardized pipelines driven by metadata enable automated, systematic data transformation, cleansing, and integration, allowing users to perform exploration, curation, and analysis tasks in GUI-based workflows.
\end{description}
\begin{figure*}[t]
	\centering
	\includegraphics[width=\textwidth]{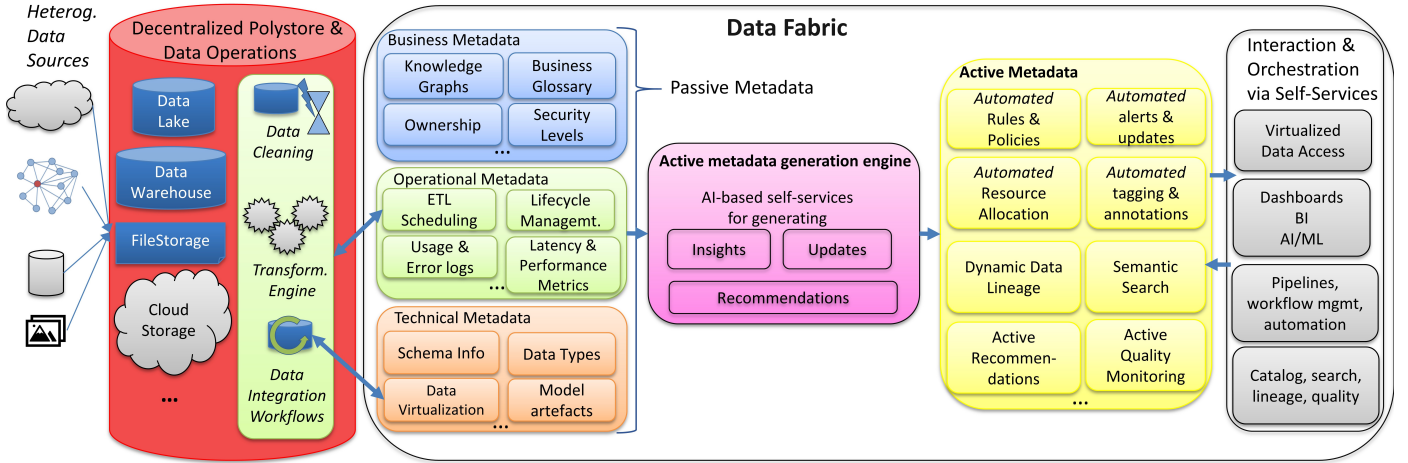}
	\caption{Data fabric architecture inspired by \cite{9671862,hechler2023data}}
	\label{fig:data_fabric_architecture}
\end{figure*}
We introduce a data fabric architecture in \cref{fig:data_fabric_architecture}, based on Hechler et al. \cite{hechler2023data}, Macias et al. \cite{macias2024data} and Priebe et al. \cite{9671862}. 
Heterogeneous data sources are first ingested into storage. 
We illustrate how the data fabric architecture utilizes various storage technologies in a decentralized polystore. 
Those may include entire data lakes, warehouses, file storages like HDFS or cloud storages like S3. 
As the fabric's main concern is not the storage itself, but the associated metadata for merging the various parts into a unified view, we decided to leave this part out of the core data fabric area. 
Similarly, we view the transformation layer for data processing at scale as belonging to the storage component. 
The fabric is responsible for coordinating the required operations. However, it is ultimately the underlying storage, that is responsible to execute resource-intensive data processing operations. 
This opposes Priebe et al. \cite{9671862}, who include storage and integration and interoperability into the fabric's core functionality. Even in their illustration one can make a case to divide a data fabric into a storage \&  operations component and a knowledge catalog section.
Essentially, we view the data fabric as knowledge catalog that connects various underlying data storages through comprehensive metadata management  capable of data access and transformation via predefined standardized data pipelines.

Following the illustration for active metadata by Hechler et al. \cite{hechler2023data}, for the data fabric we first deal with passive metadata divided into the three categories for metadata introduced by \cite{Diamantini2018}. 
Business metadata serializes the KGs and business or domain related knowledge. 
The operational metadata orchestrates ETL processes and other data processing operations on the underlying storage, collects logs and manage the data lifecycle. 
Technical metadata extracts schema information, data types, etc. Therebym it enables data virtualization and collects metadata about the connections between analytical models and their artefacts and training datasets.

These are then converted into active metadata by the responsible generation engine. 
Those include tools powered by AI methods to enable users to discover insights, perform updates on the data, and receive recommendations. 
These AI methods are not meant to be crafted by users, but rather the data fabric should provide ready-to-use self-services that are sufficiently abstracted and simplified to make anyone capable of using them. 
The self-services enable a range of data discovery methods, such as semantic search, recommendations, automated semantic labeling, tagging, and annotations. 

Finally, the interaction layer needs to provide an intuitive GUI that enables unfamiliar users to operate and utilize the self-services for integrating, searching and transforming data and orchestrating workflows.

\subsection{Data Mesh}\label{section:data_mesh}
Centralized data teams, while highly specialized, often lack business domain expertise, creating bottlenecks in scaling data-driven decision-making across the enterprise \cite{butte2022enterprise,blohm2024data}.
As different business units generate data in a decentralized manner, traditional architectures managed by central IT departments struggle to manage increasing volume and variety \cite{Draining_the_Data_Swamp,dehghani2022data,Dehghani2019}.
On the other hand, if the organization is divided in multiple business units to manage datasets specific to their domain, a lack of transparency and communication can lead to replicated effort and data silos. Here, a self-serve platform and employing proper standards through federated governance may enable reuse of datasets by others.
Furthermore, centralized data management raises concerns over data ownership and governance. Without clear responsibility for data quality and maintenance, organizations risk compromised data integrity. 
Shifting ownership back to the domains creates a strong sense of responsibility over the created data that will improve data quality \cite{dehghani2022data,bode2024towards}. 

For such reasons, Dehghani \cite{Dehghani2019,dehghani2022data} introduces the so-called data mesh, which is not (yet) a typical data architecture with concrete implementations, or infrastructure specifications, but rather ``a socio-technical data approach'' \cite{dehghani2022data}.
Interestingly, the concept draws inspiration from microservice architectures in software engineering \cite{10278300}. 
While traditional data platforms remain centralized, a data mesh applies the microservice paradigm to data architectures, where each domain functions as an independent service. Just as microservices emerged to overcome the limitations of monolithic applications, the data mesh addresses the constraints of centralized data structures. 

Another major motivating factor is the historical focus of data and analytics research on technological components rather than usage-driven approaches \cite{abbasi2016big,strengholt2023data}, which led to overlooking the significance of federated architectures, governance, and organizational design \cite{blohm2024data}. 
This is the main reason for bottlenecks in centralized data platforms. Most studies have centered on designing technical solutions rather than addressing the organizational, procedural and governance challenges inherent in large-scale data management. 
Araujo et al. \cite{DBLP:conf/centeris/Machado0S21} even go as far as calling the data mesh a paradigm shift in the field of data architectures. 
In this new paradigm, data is seen as the main concern of a truly data-oriented organization, and the pipelining tools and specific architectures itself are seen as a secondary concern. 

Several publications mention a close relationship between the data fabric and the data mesh \cite{hechler2023data,strengholt2023data,blohm2024data,serra2024deciphering,bode2024towards}.
James Serra \cite{serra2024deciphering} describes data fabrics as an architectural and technological framework that operates within or across domains in a data mesh, supporting its components and enabling synergy rather than competition. While a data mesh offers a decentralized, domain-oriented blueprint, data fabrics provide the underlying infrastructure to implement it. Multiple data fabrics can coexist within a mesh, and this integration approach also extends to other data architectures like data warehouses or lakehouses.

Despite the theoretical foundation of a data mesh and related frameworks, research on their real-world application remains limited.
While studies have explored technological architectures, privacy concerns, and individual case studies, there is little systematic analysis of how industry implements these concepts \cite{bode2024towards}.
However, Araujo et al. \cite{DBLP:conf/centeris/Machado0S21} outline two industrial use cases:

Zalando \cite{Schultze2020}, a leading European fashion platform, initially relied on a centralized data lake but faced challenges such as lack of data ownership, poor data quality, and scalability bottlenecks as data sources and consumers grew.
To address these issues, the company implemented a data mesh, introducing decentralized data ownership, domain-oriented teams, and a shift toward treating data as a product rather than a by-product. 
This transition enabled interoperability and decentralizing data responsibilities and improved governance through a metadata layer.

Netflix \cite{Cunningham2020,Pires2023,Nguonly2021}, serving over 150 million global users, generates petabytes of data daily and sought to integrate its studios into a unified system capable of handling this massive volume efficiently.
Due to redundant pipeline efforts, maintenance overhead, lack of best practices, high latency, and error correction issues, Netflix implemented a data mesh that abstracts pipeline complexity for users. 
The infrastructure allows users to build pipelines while accessing a metadata catalog and standardized processes without needing deep technical knowledge.


\subsubsection{Characteristics}
Four principles define the data mesh as originally introduced by Dehghani \cite{Dehghani2019}: 
\begin{description}
	\item[DM1 ---]\phantomsection\label{DM1}  \textbf{Data-as-a-product}: creation, management, and distribution of data products via a catalog. 
    Data providers are responsible for publishing and provisioning them and collaborate with data owners and stewards for maintenance.

	\item[DM2 ---]\phantomsection\label{DM2}  \textbf{Decentralized data ownership and architecture}: each domain has ownership over its data products, ensuring that domain teams are responsible for the quality, maintenance, and lifecycle of their data products. 

	\item[DM3 ---]\phantomsection\label{DM3}  \textbf{Self-Service capabilities}: a self-service data platform must be established to enable domain teams to create, manage, and consume data products without requiring extensive technical expertise. 

	\item[DM4 ---]\phantomsection\label{DM4}  \textbf{Federated governance}: a federated governance model balances decentralized autonomy with global governance standards, ensuring that policies are enforced consistently across all domains while allowing flexibility for domain-specific rules. 
\end{description}
The data-as-a-product principle frames data products as deliberately distributed and interconnected across individual ''mesh nodes'' \cite{DBLP:conf/centeris/Machado0S21,soininen2025data}.
To give an example for a data product, consider the \textit{MLFlow} model export functionality that outputs not only the trained model parameters, but also essential metadata, dependencies, and governance information. 
Each exported model includes environment specifications such as a \texttt{requirements.txt} or \texttt{conda.yaml}, ensuring reproducibility across different infrastructure setups. 
Additionally, \textit{MLFlow} tracks experiment metadata, including training hyperparameters, dataset versions, and performance metrics, allowing for model lineage and governance.  
 
Eichler et al. \cite{DBLP:conf/summersoc/EichlerGHSM22} recognize that data providers often lack incentives to share their data, because they face additional effort without immediate benefits, discouraging participation. 
To address this, they subdivide data products further into data assets. 
Data assets are raw, registered data in a company’s catalog, while data products are refined versions with defined metadata, usage policies, and provisioning options. 
The transformation process follows three scenarios: (1) a provider directly registers data-as-a-product, (2) a consumer requests access to a data asset, prompting its transformation, or (3) a steward or another employee enriches metadata and submits it for approval. 
This staged approach reduces the initial burden on providers, ensuring they only need to add metadata when data is actually requested. 

Domain-oriented ownership shifts data responsibility to individual business domains, allowing them to manage and utilize their own data according to their specific operational logic.
This model fosters agility and autonomy by structuring data management around product-focused teams \cite{DBLP:conf/centeris/Machado0S21,bode2024towards}).

Self-service platforms serve as a technical abstraction layer, enabling unified access to data products while reducing complexity for users \cite{10248262,bode2024towards}. 
By providing a standardized infrastructure, they eliminate redundant technical efforts and simplify the management of data products across their lifecycle \cite{dehghani2022data,bode2024towards,araujo2022advancing}.


\begin{figure*}[t]
		\centering
		\includegraphics[width=0.9\textwidth]{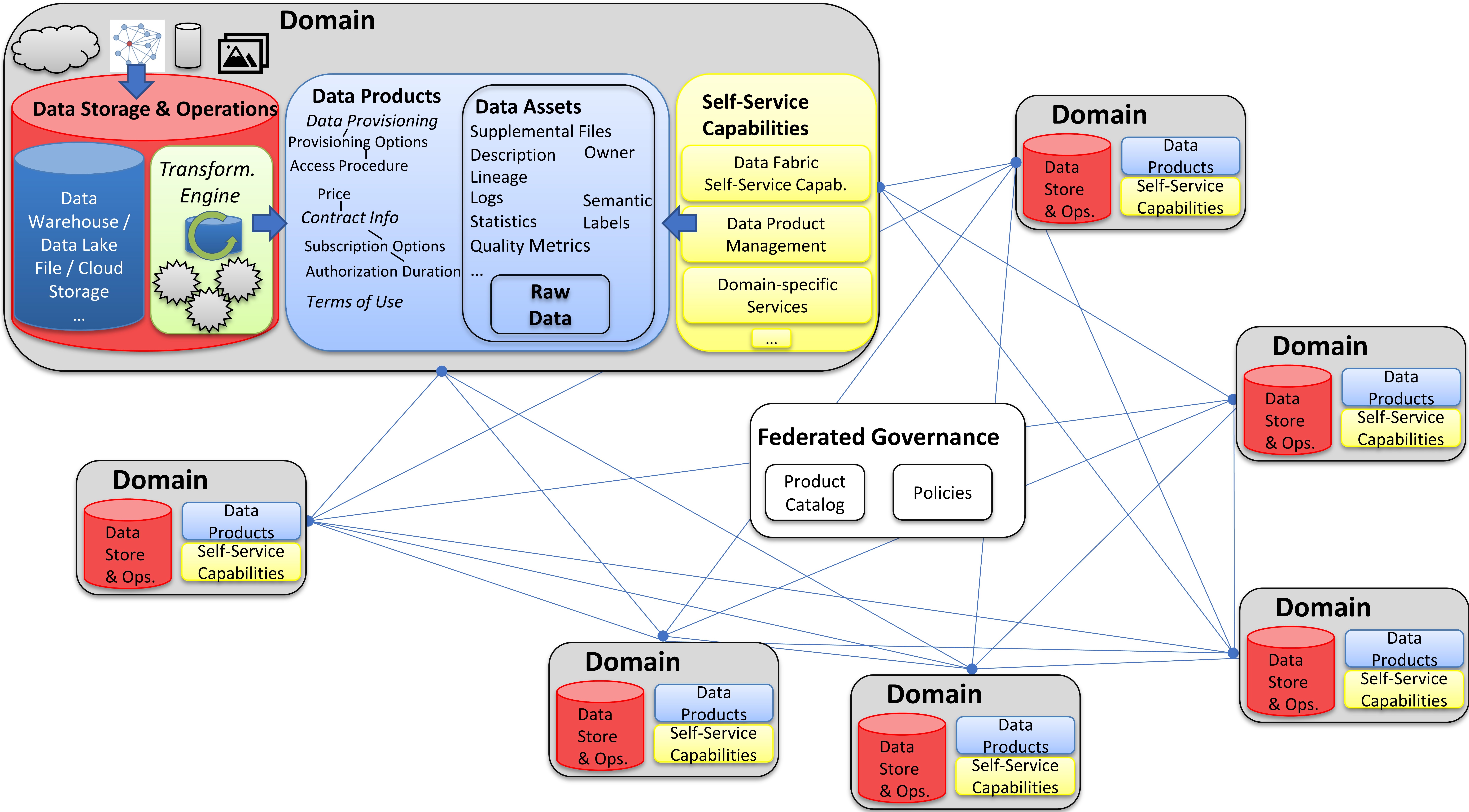}
		\caption{Data mesh architecture inspired by \cite{DBLP:conf/summersoc/EichlerGHSM22,dolhopolov2024implementing,hechler2023data,strengholt2023data}}
		\label{fig:datamesh}
\end{figure*} 

Finally, federated computational governance establishes global governance standards that ensure interoperability and secure data access while enabling decentralized and autonomous governance by distributed teams. 
By balancing centralized standards with decentralized decision-making, federated governance ensures that data remains accessible and governed in a structured yet flexible manner \cite{dehghani2022data,bode2024towards,10248262}.
Policies are also designed to balance local, i.e. domain-oriented, autonomy with global standardization.
The federated governance model oversees security policies (defining authentication, authorization, and access control), interoperability policies (standardizing schema definitions, data contracts, and API structures), communication policies (guidelines for cross-domain data sharing), and documentation policies (requiring metadata enrichment, lineage tracking, and accessibility compliance)

For a definition we follow Blohm et al. \cite{blohm2024data} once more:
\begin{definition}\label{definition-datamesh}[\textbf{Data Mesh}]
	``\textit{Data mesh is a socio-technical, decentralized, distributed concept for enterprise data management \cite{bode2024towards}. It is characterized by the four principles of data-as-a-product, domain orientation, self-service platforms, and federated computational governance \cite{Dehghani2020}.''}
\end{definition}

\Cref{fig:datamesh} illustrates the data mesh architecture. 
Van der Werf et al. \cite{van2024towards} identify four archetypes, out of which the fine-grained fully federated mesh or short pure data mesh is considered to be the most theoretical and decentralized but also the most mature version. 
We have chosen this pure form in which individual nodes directly exchange data products and deliberately left out a central distribution layer. 
Every atomic team or business unit is responsible for a single domain and the data products arising from it. 
While this approach provides high flexibility and strong domain specialization, it might also introduce challenges such as wasting resources, when data transformations are performed redundantly due to lack of communication or transparency between domains.  
Here, the maturity of the self-service capability and federated governance components are particularly relevant.
In the figure, one node is enlarged representatively to display more details.
In each domain heterogeneous raw data emerges that will be stored in a data storage and operations platform of choice. 
Those can be whatever is available and suitable in the specific domain, e.g. data warehouse/lake, cloud storages or file storages etc. 
Each domain is responsible for transforming and packaging raw data into data products. 
Data products are further subdivided into data assets and raw data as inspired by Eichler et al. \cite{DBLP:conf/summersoc/EichlerGHSM22}. 
In each domain, data consumers and providers have access to self-service capabilities. In the figure, the individual nodes inherit self-service capabilities from a data fabric to represent the close relationship and potential synergy between the two paradigms. 
They are extended by domain-specific services as well as data product management where domain teams register their data products along with associated metadata, quality metrics, and access policies and facilitate the distribution to other domains. 
This data product management and distribution is overseen by a central data federated governance layer to enforce global governance standards consistently ensuring policies, interoperability and secure data access.

\section{Evaluation}\label{section:evaluation}
The evaluation begins with a definition of the criteria used to assess and describe the discussed architectures. 
The criteria are defined in a way that allows to assign a quantitative measure (see \cref{tab:evaluation_dimensions}).
Finally, these evaluation criteria in conjunction with the technical requirements from \cref{section:data_architectures} are evaluated against real-world systems to find a suitable match.

\subsection{Dimensional Analysis}\label{section:dater_dimensions}
The preceding literature review and the in-depth examination of the discussed data architectures helps us to derive the following dimensions.
These aim to (a) describe the characteristics of a given architecture, (b) highlight similarities and differences between them, and (c) provide a framework for assigning existing systems to one of the six categories in \cref{section:data_architectures}:
\paragraph{Control, Governance \& Trust}
\begin{description}
    \item[D1 – Architectural Paradigm:]\phantomsection\label{D1} Measures how centralized or decentralized the data architecture is in terms of control, storage infrastructure, and processing. Higher scores reflect distributed, domain-driven models that promote autonomy.
    
    \item[D2 – Governance \& Ownership:]\phantomsection\label{D2} Examines the model of data control from centralized to federated ownership. Higher values indicate domain-level accountability and governance.
    
    \item[D3 -  Security \& Trust]\phantomsection\label{D3} Measures how much the architecture gives priority to support secure access, data protection, auditability, and trust-building mechanisms such as data provenance, policy enforcement, or verifiable sharing.
\end{description}
\paragraph{Data Modeling \& Understanding}
\begin{description}
    \item[D4 – Data Formats:]\phantomsection\label{D4} Evaluates how much the architecture considers flexibility in handling different data formats. Higher scores indicate native support for polystores and unstructured, semi-structured, and structured data.

    \item[D5 – Metadata Management:]\phantomsection\label{D5} Assesses how metadata is handled from manual and static to dynamic, automated, and lineage-aware systems. Advanced setups foresee the use of AI or active metadata.

    \item[D6 – Knowledge Management:]\phantomsection\label{D6} Assesses the extent to which the architecture supports modeling, managing, or leveraging domain knowledge alongside data. 
    This includes formal approaches (e.g., SDM) and informal ones, such as business glossaries, tagging systems, or e.g., enterprise LLM-based chatbots \citep{yun2025eicopilot} with  richer data understanding, reasoning, and integration.
\end{description}
\paragraph{Data Processing}
\begin{description}
    \item[D7 – Integration Focus:]\phantomsection\label{D7} Captures how central integration is to the architecture’s design. A high score means data integration is a foundational concern rather than an afterthought.
    
    \item[D8 – Data Virtualization:]\phantomsection\label{D8} Describes how much focus the architecture places on abstracting storage details to enable querying flexibly and transparently without having to move data physically. 
    
    \item[D9 – ML \& Analytics Support:]\phantomsection\label{D9} Captures the degree of support for ML and data science. Higher scores indicate built-in tools and pipelines for advanced analytics beyond traditional BI.
\end{description}

\begin{table*}[!htbp]
    \centering
    \scriptsize
    \begin{tabularx}{\textwidth}{|p{0.35cm}|L{2.5cm}|L{2.2cm}|X|X|}
        \hline
        \textbf{\#} & \textbf{Dimension} & \textbf{1 = Low} & \textbf{3 = Medium} & \textbf{5 = High} \\
        \hline 
        
        \multicolumn{5}{|c|}{\textbf{Control, Governance \& Trust}} \\[2pt] 
        \hline
        
        \hyperref[D1]{\textbf{D1}} & Architectural Paradigm & Centralized & Federated & Decentralized \\ \hline
        
        \hyperref[D2]{\textbf{D2}} & Governance \& Ownership & Centralized & Shared or hybrid & Federated / Domain-oriented \\ \hline
        
        \hyperref[D3]{\textbf{D3}} & Security \& Trust & Minimal/static security & Moderate control with some auditing & Integrated, policy-enforced, trust-aware \\
        \hline 
        
        \multicolumn{5}{|c|}{\textbf{Data Modeling \& Understanding}} \\[2pt] 
        \hline
        
        \hyperref[D4]{\textbf{D4}} & Data Formats & Structured only & Multi-format, some flexibility & Polystore / All formats \\ \hline
        
        \hyperref[D5]{\textbf{D5}} & Metadata Management & Manual / fixed & Automated extraction / tagging & AI-driven, dynamic, lineage-aware \\ \hline
        
        \hyperref[D6]{\textbf{D6}} & Knowledge Management & None & Basic domain modeling (glossaries, tags, rules) & Integrated knowledge layer (ontologies/KGs, LLMs) \\
        \hline 
        
        \multicolumn{5}{|c|}{\textbf{Data Processing}} \\[2pt]
        \hline
        
        \hyperref[D7]{\textbf{D7}} & Integration Focus & Not central to the architecture & Localized or externalized & Core architectural concern \\ \hline
        
        \hyperref[D8]{\textbf{D8}} & Query Mechanisms & Local SQL only & SQL + NoSQL / partial federation & Federated / virtualized querying \\ \hline
        
        \hyperref[D9]{\textbf{D9}} & ML \& Analytics Support & BI only & Some support for ML & Full ML pipeline and data science support \\ \hline
        
    \end{tabularx}
    \caption{Evaluation Dimensions for Data Architectures on a scale from 1 to 5, mapped to General Requirements}
    \label{tab:evaluation_dimensions}
\end{table*}

\renewcommand{\arraystretch}{1.5}
\setlength{\tabcolsep}{4pt}
\begin{table*}[!htbp]
    \centering
    \small
    \resizebox{0.75\textwidth}{!}{%
    \begin{tabular}{|c|L{2cm}|L{2cm}|L{2.5cm}|L{2cm}|L{2cm}|L{2.2cm}|}
        \hline
         & \textbf{Data Warehouse \citep{quix2003metadatenverwaltung,inmon2005building,bose2009advanced}} & \textbf{Data Lake \citep{hai2023data,Quix2019,azzabi2024data,10.14778/3352063.3352116}} & \textbf{Semantic Data Lake \citep{hoseini2023semantic,Dibowski2020a,Bagozi2019}} & \textbf{Data Lakehouse \citep{10020719,schneider2024lakehouse,armbrust2021lakehouse}} & \textbf{Data Fabric \citep{hechler2023data,blohm2024data,serra2024deciphering}} & 
         \textbf{Data Mesh \citep{dehghani2022data,serra2024deciphering,DBLP:conf/centeris/Machado0S21}} \\
        \hline 
        \multicolumn{7}{|c|}{\textbf{Control, Governance \& Trust}} \\[2pt] 
        \hline
        \hyperref[D1]{\textbf{D1}} & 
        1 –  Centralized relational &
        1 –  Centralized raw storage &
        1 –  Semantic layer on centralized DL &
        1 –  Hybrid of DW \& DL &
        3 –  Federated integration \& Active Metadata &
        5 –  Decentralized domain–oriented \\
        \hline
        \hyperref[D2]{\textbf{D2}} & 
        1 –  Centralized &
        1 –  Centralized &
        1 –  Centralized &
        1 –  Centralized &
        3 –  Policy–managed &
        5 –  Domain ownership \\
        \hline
        \hyperref[D3]{\textbf{D3}} & 
        2 –  Strong role–based security but typically centralized and static &
        1 –  Security is often minimal or bolted–on; access control at storage layer &
        2 –  semantic technologies to account for it, but not a primary focus &
        3 –  security models via modern engines, but not the primary focus &
        5 –  central &
        4 –  Domain–level responsibility; requires strong governance frameworks to scale \\
        \hline 
        \multicolumn{7}{|c|}{\textbf{Data Modeling \& Understanding}} \\[2pt] 
        \hline
        \hyperref[D4]{\textbf{D4}} & 
        1 –  relational &
        5 –  raw data in original format &
        5 –  raw data in original format &
        3 –  Mainly relational &
        5 –  arbitrary format &
        5 –  Data products of arbitrary format \\
        \hline
        \hyperref[D5]{\textbf{D5}} & 
        1 –  Fixed, scoped model &
        3 –  Automated metadata extraction &
        4 –  Metadata enriched with semantics &
        3 –  Automated metadata extraction &
        5 –  AI–driven, automated &
        2 –  Decentralized, on–demand \\
        \hline
        \hyperref[D6]{\textbf{D6}} & 
        1 –  None &
        1 –  None &
        5 –  Ontologies \& KGs &
        3 –  Basic &
        5 –  KG–based metadata &
        2 –  local vocabularies, but no global semantic layer \\
        \hline 
        \multicolumn{7}{|c|}{\textbf{Data Processing}} \\[2pt]
        \hline
        \hyperref[D7]{\textbf{D7}} & 
        5 –  ETL &
        3 –  on–demand integration (ELT) &
        4 –  Semantic–level integration &
        2 –  Mix between DL and DW &
        4 –  Integration driven by metadata  &
        2 –  Local integration within data products \\
        \hline
        \hyperref[D8]{\textbf{D8}} & 
        2  –  access remains tightly coupled to internal engine and physical schemas &
        3  –  Raw file-based querying; flexibility via tools like Spark/Presto, but limited abstraction &
        5 –  OBDA-based access over heterogeneous data using semantic models. &
        3 –  Combines files and tables with SQL engines, but limited abstraction from physical storage &
        5 –  virtualized access, logical layers, and cross-platform queries &
        3 –  Varies by domain; considers APIs or connectors; lacks global query abstraction \\
        \hline
        \hyperref[D9]{\textbf{D9}} & 
        1 –  Focus on reporting \& BI &
        2 –  Not the primary focus &
        1 –  No focus &
        5 –  Full ML lifecycle support  &
        3 –  Moderate &
        1 — Not a focus \\
        \hline
    \end{tabular}}
    \caption{Evaluation of Data Architectures against requirements
    }
    \label{tab:data_architecture_evaluation}
\end{table*}

\subsection{Evaluation of Real-World Systems}\label{section:dater_real_world_systems}
\paragraph{The Semantic Data Reservoir} (\fcolorbox{blue!20}{white}{\href{https://github.com/hsnr-data-science/SEDAR}{SEDAR}}) \cite{hoseini2023sedar}
has been designed as a modular system using open-source big data technologies, making it suitable as a workbench for research on scalable data architectures. 
Originally a classical data lake, SEDAR has evolved to support semantic data management (SDM) and ontology-based data access (OBDA) \cite{Paulus2021,hoseini2023semantic}.
SEDAR streamlines the ML life cycle using MLOps principles \cite{alla2020mlops}, supports AutoML \cite{hoseini2024enhancing} and is applied in production as part of a cyber-physical system for an Industry 4.0 case \cite{hoseini2024coatings}.
In the latest progress, a novel LLM-driven natural language interface has been added, which is designed to complement the user interface and to lower the technical barrier for users.
The chat-based interface enables users to perform complex data management tasks using everyday language by leveraging the ability of LLMs to convert user queries into actions executed on the system.

From a control and governance perspective \hyperref[D1]{(D1–D3)}, SEDAR initially exhibits centralized characteristics. 
Security and policy enforcement remain basic.
In terms of data modeling and understanding \hyperref[D4]{(D4–D6)}, SEDAR supports heterogeneous formats, semantic enrichment, and the use of ontologies and knowledge graphs as first-class citizens, making it highly suitable for knowledge-driven applications. 
Metadata management is automated and partially AI-augmented.
Data processing \hyperref[D7]{(D7-D9)} in SEDAR treats integration as a foundational concern, leverages Apache Spark for scalable processing, supports OBDA, and implements the full ML lifecycle following MLOps principles. 

SEDAR partially overlaps with data lakehouse and warehouse paradigms due to its support for relational processing, SQL-based querying, and ML workflows.
However, it does not enforce schema-on-write \hyperref[DW1]{DW1}  and the same storage type and data format across sources \hyperref[DLH1]{DLH1} and does not prioritize OLAP-centric optimization \hyperref[DW7]{DW7}.
Given its semantic foundation, integration-centric design, and AI-enhanced capabilities, SEDAR increasingly fulfills key \textbf{data fabric} criteria \hyperref[DF2]{(DF2-DF4)}, although currently SEDAR implements a centralized storage layer \hyperref[DF1]{DF1}. 
Hence, it moves away from the scope of classical data lakes, warehouses, and lakehouses by offering semantic interoperability, self-service capabilities and advanced automation across the data lifecycle.

\paragraph{The Stackable Data Platform} Stackable\footnote{\url{https://stackable.tech/}, accessed on 02.05.2026} is an open-source modular data platform built on Kubernetes that integrates a curated set of big data and data science tools into a cohesive platform-as-code architecture. It provides production-ready deployments of components like Kafka, Trino, Open Policy Agent, Hive via Kubernetes operators, with strong support for declarative configuration and infrastructure management.

From a control and governance perspective \hyperref[D1]{(D1–D3)}, Stackable promotes a hybrid governance model. 
While orchestration and policy enforcement can be centralized (e.g., via Kubernetes and Open Policy Agent), the platform enables domain-level autonomy in configuring and managing services, fostering decentralized ownership. 
It supports advanced access control through integration offering a relatively mature security and trust enforcement model.
In terms of data modeling and understanding \hyperref[D4]{(D4–D6)}, Stackable is tool-agnostic and does not prescribe a specific data model.
It supports diverse data types through connectors and engines, but leaves semantic enrichment and metadata management to external services. 
Similarly, ontologies and knowledge graphs are not natively integrated but can be layered in through additional tooling. 
Regarding data processing \hyperref[D7]{(D7–D9)}, Stackable enables large-scale data integration and analytics through its orchestration of distributed compute engines like Spark and Flink, and supports virtualization and query federation via Trino. 
However, it lacks global semantic abstraction and unified orchestration of ML workflows. While the platform supports full ML pipelines via integrations, MLOps is not a primary design focus.

Stackable does not conform strictly to data lakehouse or data fabric paradigms due to its lack of unified metadata, active governance layers, or enforced semantic models. 
However, it aligns most closely with \textbf{data mesh} principles particularly through its domain-oriented deployment patterns \hyperref[DM2]{(DM2)}, emphasis on infrastructure-as-code, and federated governance \hyperref[DM4]{(DM4)}. 
While it lacks the organizational enforcement, its architectural modularity and governance extensibility make it a strong enabler for infrastructure-centric data mesh implementations.
If these are extended by self-service capabilities \hyperref[DM3]{(DM3)} in the future, possibly by integrating corresponding external services, then Stackable covers a mature technological base for data meshes.

\paragraph{Microsoft Fabric} Microsoft Fabric\footnote{\url{https://www.microsoft.com/en-us/microsoft-fabric}, accessed on 02.05.2026} is Microsoft’s platform-as-a-service offering, unifying components from Power BI, Azure Synapse, Azure Data Factory, and OneLake.
Designed for end-to-end data and AI workloads, it provides services for data engineering, data science, real-time analytics, and business intelligence — with strong emphasis on cloud-native scalability, collaboration, and ease of use.

From a control and governance perspective \hyperref[D1]{(D1–D3)}, Microsoft Fabric implements a federated orchestration model layered over centralized infrastructure. While the platform itself is centrally managed (e.g., via Azure), governance can be delegated across workspaces and domains. 
It offers comprehensive role based access management (RBAC), lineage tracking, and sensitivity labeling features.
Data access and governance is handled by Microsoft Purview, though policy-based data sharing is still evolving.
In terms of data modeling and understanding \hyperref[D4]{(D4–D6)}, Fabric supports a wide range of data formats promoting a polystore architecture via OneLake, where all data can be virtually accessed through a unified namespace. 
Metadata extraction and enrichment are automated and AI-augmented, particularly through Microsoft Copilot. 
While domain modeling support exists (via data cataloging, tagging, glossaries), deeper semantic constructs such as ontologies or knowledge graphs are not first-class features.
Regarding data processing \hyperref[D7]{(D7–D9)}, Fabric offers highly integrated support for ETL pipelines, real-time streaming, and analytical processing.
It supports distributed querying across files and tables, but semantic virtualization and ontology-based abstraction are not core design elements. 
Fabric fully integrates AutoML and MLOps capabilities enabling full ML lifecycle support from within the platform.

Microsoft Fabric aligns closely with the \textbf{data fabric} paradigm \hyperref[DF1]{(DF1, DF3 \& DF4)}, particularly in its unified infrastructure, active metadata, and orchestration of cross-role workflows. 
While data virtualization can be configured via so-called \textit{shortcuts}, it falls short of the semantic richness for advanced data virtualization \hyperref[DF2]{(DF2)}. 
Overall, Microsoft Fabric represents a commercially mature, enterprise-oriented fabric implementation that balances manageability and innovation, with strong vertical integration across Microsoft’s data ecosystem.

\begin{table*}[t]
    \centering
    \small
    \renewcommand{\arraystretch}{1.2}
    \resizebox{0.9\linewidth}{!}{%
    \begin{tabular}{|c|L{3cm}|L{5cm}|L{5cm}|}
        \hline
         & \textbf{SEDAR \citep{hoseini2023sedar}}  
         & \textbf{Stackable \citep{stackable2025interview}} 
         & \textbf{Microsoft Fabric \citep{ghosh2024mastering}} \\
        \hline 

        \multicolumn{4}{|c|}{\textbf{Control, Governance \& Trust}} \\ 
        \hline
        
        \hyperref[D1]{\textbf{D1}} & 
        1 – Centralized  & 
        5 – Decentralized &
        3 – centralized governance, but workspace-level autonomy \& logical domains \\ 
        \hline
        
        \hyperref[D2]{\textbf{D2}} & 
        1 – Centralized control plane & 
        4 – centrally policy-managed, but configurable from within domains & 
        3 – centralized governance tools, but support for role- and workspace-based ownership models \\ 
        \hline
        
        \hyperref[D3]{\textbf{D3}} & 
        1 – Basic RBAC; not a strong focus & 
        4 – sophisticated and scalable measures & 
        3 – fine-grained access-policies, but policy-based data sharing are not core.  \\ 
        \hline 

        \multicolumn{4}{|c|}{\textbf{Data Modeling \& Understanding}} \\ 
        \hline
        
        \hyperref[D4]{\textbf{D4}} & 
        3 – Raw data retained; main support relational & 
        3 – arbitrary data can be processed, but support mainly designed for relational model &
        5 – services integrate a variety of formats suitable for polystore-style needs. \\ 
        \hline
        
        \hyperref[D5]{\textbf{D5}} & 
        5 – Automatically enriched with semantics & 
        1 – Outsourced to external tooling & 
        5 – AI-augmented, consistent with modern active metadata \\ 
        \hline
        
        \hyperref[D6]{\textbf{D6}} & 
        5 – Ontologies \& Knowledge Graphs & 
        1 – None & 
        3 – basic domain modeling capabilities such as tagging, schemas, and glossaries. \\ 
        \hline 

        \multicolumn{4}{|c|}{\textbf{Data Processing}} \\
        \hline
        
        \hyperref[D7]{\textbf{D7}} & 
        5 – Integration is central & 
        2 – ingestion and ELT workflows, integration is often manual or tool-driven & 
        5 – Integration is central \\ 
        \hline
        
        \hyperref[D8]{\textbf{D8}} & 
        5 – OBDA & 
        3 – virtualization via Trino connectors, but no global query abstraction & 
        4 – can be configured via OneLake, but no global semantic layer \\ 
        \hline
        
        \hyperref[D9]{\textbf{D9}} & 
        5 – Full ML lifecycle (incl. AutoML) & 
        2 – possible, but not a special focus & 
        5 – Full ML lifecycle \\
        \hline
    \end{tabular}
    }
    \caption{Evaluation of real-world data architecture systems based on technical requirements.}
    \label{tab:systems_evaluation}
\end{table*}

\section{Related Surveys \& Work}\label{section:dater_related_work}
The ISO/IEC TR 20547-1:2020 standard \citep{iso20547-1} defines a reference architecture for big data systems, providing a high-level framework to guide the design of interoperable and scalable solutions. While DATER shares the goal of architectural classification, it places greater emphasis on comparative evaluation across diverse architectures based on concrete technical requirements.
A whitepaper by Fraunhofer ISST explores the conceptual and technical intersections between data mesh and data spaces \citep{fraunhofer2023datamesh}, highlighting their complementary nature in decentralized data ecosystems. It argues that data mesh offers organizational and governance principles within enterprises, while data spaces provide the infrastructure for secure, cross-organizational data sharing.
\cite{9671862} present architecture diagrams in ArchiMate notation \citep{josey2019archimate} for data warehouse, data mesh, and data fabric based on \citep{international2017dama}.
\cite{atzori2024dataspaces} compare data spaces to data warehouses, lakes and different types of database systems.

In his book \citep{serra2024deciphering}, James Serra gives a high-level introduction into modern data architectures: data warehouse, lake, lakehouse, fabric and mesh. As a textbook, it is more oriented towards readers unfamiliar with the field, explaining many concepts in much detail, which we assume to be available with readers of this article.
We draw inspiration from this work heavily but shift the focus towards more in-depth technical detail, comparisons, and add more architectures.

A systematic literature review by Ataei and Litchfield \citep{ataei2022state} reveals that many data architectures rely on monolithic data pipelines with central storage, which hinders scalability and maintainability. 
Moreover, they point out that most existing  reference architectures align closely with traditional data warehousing concepts, often integrating data lakes in a way that does not fully address modern data processing challenges.
A major conclusion from their work is the necessity for decentralized and distributed big data architectures suggesting that emerging paradigms such as data mesh could better support scalability, data ownership, and interoperability  as compared to centralized architectures. 
However, they note that industry adoption of such architectures remains slow due to the high cost and complexity of implementation.
Additionally, the study identifies a lack of rigorous evaluation methodologies, making it difficult to assess their effectiveness in practice. 
Our work builds upon these insights by providing a more in-depth technical comparison of modern data architectures. 
\section{Conclusion}\label{section:dater_conclusion}
In response to the growing complexity of modern data landscapes, this chapter provides a structured exploration and comparison of contemporary data architectures. 

The study reveals that no single architecture universally satisfies all technical needs, underscoring the importance of context-aware selection and hybrid solutions.
This led to convergence of architectural paradigms, e.g., lakehouses and fabrics blending the strengths of lakes and warehouses. 
This comes with a lack of consensus in the academic and industrial literature regarding the definitions and boundaries of and between certain architectures, indicating a need for a clear characterization of data architectures that motivated this work.

Modern data ecosystems increasingly shift toward decentralization, which promote domain-oriented ownership, federated governance, and peer-to-peer data sharing.
The increasing importance of metadata and semantic technologies across architectures, highlights another shift toward more intelligent, discoverable, and machine-interpretable data systems.
Architectures like data mesh and data fabric reflect more than technical designs. 
They also embody operational principles, organizational models, and mark a third cultural shift in data management.

\section*{Funding}
This work was supported by the German Federal Ministry of Education and Research (BMBF) under grant no.~13FH557KX0 (project i2DACH).


\section*{Declaration of Generative AI and AI-assisted technologies in the writing process}
During the preparation of this work, the authors used ChatGPT, DeepL, and Grammarly to improve phrasing and clarity. After using these tools, the authors reviewed and edited the content and take full responsibility for the final version.

\bibliographystyle{elsarticle-num}
\bibliography{references}

@article{dolhopolov2024implementing,
  title={Implementing Federated Governance in Data Mesh Architecture},
  author={Dolhopolov, Anton and Castelltort, Arnaud and Laurent, Anne},
  journal={Future Internet},
  volume={16},
  number={4},
  pages={115},
  year={2024},
  publisher={MDPI}
}

@inproceedings{van2024towards,
  title={Towards a Data Mesh Reference Architecture},
  author={van der Werf, Daniel and Moreira, Jo{\~a}o and Piest, Jean Paul Sebastian},
  booktitle={International Conference on Enterprise Design, Operations, and Computing},
  pages={339--353},
  year={2024},
  organization={Springer}
}

@article{ataei2022state,
  title={The state of big data reference architectures: A systematic literature review},
  author={Ataei, Pouya and Litchfield, Alan},
  journal={IEEE Access},
  volume={10},
  pages={113789--113807},
  year={2022},
  publisher={IEEE}
}

@book{josey2019archimate,
  title={ArchiMate{\textregistered} 3.1-A Pocket Guide},
  author={Josey, Andrew},
  year={2019},
  publisher={Van Haren}
}

@misc{Dehghani2020,
  author       = {Zhamak Dehghani},
  title        = {Data Mesh Principles and Logical Architecture},
  year         = {2020},
  url          = {https://martinfowler.com/articles/data-mesh-principles.html},
  note         = {Accessed: 18 Jan 2024}
}

@misc{fraunhofer2023datamesh,
  author       = {Nils Jahnke and Katrin Bendiek and Malina Klue\ss},
  title        = {{Data Mesh und Data Spaces – Zwei Konzepte für dezentrale Datenökosysteme}},
  year         = {2023},
  howpublished = {Whitepaper Fraunhofer ISST, \url{https://www.bk.admin.ch/dam/bk/de/dokumente/dti/DatenoekosystemSchweiz/Grundlagen/Fraunhofer-ISST_Data-Mesh-und-Data-Spaces_Whitepaper%20.pdf.download.pdf/Fraunhofer-ISST_Data-Mesh-und-Data-Spaces_Whitepaper%20.pdf}},
  note         = {Accessed: 2025-06-02}
}

@book{strengholt2023data,
  title={Data Management at Scale},
  author={Strengholt, Piethein},
  year={2023},
  publisher={" O'Reilly Media, Inc."}
}

@misc{iso20547-1,
  title        = {{ISO/IEC TR 20547-1:2020: Information technology — Big data reference architecture — Part 1: Framework and application process}},
  author       = {{International Organization for Standardization}},
  year         = {2020},
  howpublished = {\url{https://www.iso.org/standard/71275.html}},
  note         = {Accessed: 2025-06-02}
}

@inproceedings{DBLP:conf/summersoc/EichlerGHSM22,
  author       = {Rebecca Eichler and
                  Christoph Gr{\"{o}}ger and
                  Eva Hoos and
                  Holger Schwarz and
                  Bernhard Mitschang},
  editor       = {Johanna Barzen and
                  Frank Leymann and
                  Schahram Dustdar},
  title        = {From Data Asset to Data Product - The Role of the Data Provider in
                  the Enterprise Data Marketplace},
  booktitle    = {Service-Oriented Computing - 16th Symposium and Summer School, SummerSOC
                  2022, Hersonissos, Crete, Greece, July 3-9, 2022, Revised Selected
                  Papers},
  series       = {Communications in Computer and Information Science},
  volume       = {1603},
  pages        = {119--138},
  publisher    = {Springer},
  year         = {2022},
  url          = {https://doi.org/10.1007/978-3-031-18304-1\_7},
  doi          = {10.1007/978-3-031-18304-1\_7},
  bibsource    = {dblp computer science bibliography, https://dblp.org}
}

@misc{Schultze2020,
  author    = {Max Schultze and Arif Wider},
  title     = {Data Mesh in Practice: How Europe’s Leading Online Platform for Fashion Goes Beyond the Data Lake},
  year      = {2020},
  howpublished = {Online video},
  note      = {Available at: \url{https://www.youtube.com/watch?v=eiUhV56uVUc} [Accessed 06.02.205]}
}

@INPROCEEDINGS{10248262,
  author={Wider, Arif and Verma, Sumedha and Akhtar, Atif},
  booktitle={2023 IEEE International Conference on Web Services (ICWS)}, 
  title={Decentralized Data Governance as Part of a Data Mesh Platform: Concepts and Approaches}, 
  year={2023},
  volume={},
  number={},
  pages={746-754},
  doi={10.1109/ICWS60048.2023.00101}}

@misc{Cunningham2020,
  author    = {Justin Cunningham},
  title     = {Netflix Data Mesh: Composable Data Processing - Justin Cunningham},
  year      = {2020},
  howpublished = {Online video},
  note      = {Available at: \url{https://www.youtube.com/watch?v=TO_IiN06jJ4} [Accessed 06.02.205]}
}

@misc{Pires2023,
  author       = {Guil Pires and Mark Cho and Mingliang Liu and Sujay Jain},
  title        = {Streaming SQL in Data Mesh},
  year         = {2023},
  month        = {November},
  url          = {https://netflixtechblog.com/streaming-sql-in-data-mesh-0d83f5a00d08},
  note         = {Accessed: 2025-02-06},
  howpublished = {Netflix TechBlog}
}

@misc{Nguonly2021,
  author       = {Andrew Nguonly and Armando Magalhães and Obi-Ike Nwoke and Shervin Afshar and Sreyashi Das and Tongliang Liu and Wei Liu and Yucheng Zeng},
  title        = {Data Movement in Netflix Studio via Data Mesh},
  year         = {2021},
  month        = {July},
  url          = {https://netflixtechblog.com/data-movement-in-netflix-studio-via-data-mesh-3fddcceb1059},
  note         = {Accessed: 2025-02-06},
  howpublished = {Netflix TechBlog}
}

@book{ghosh2024mastering,
  title={Mastering Microsoft Fabric},
  author={Ghosh, Debananda},
  year={2024},
  publisher={Springer}
}

@misc{stackable2025interview,
  author       = {Stackable Data Platform Team},
  title        = {Interview on the Stackable Data Platform},
  year         = {2025},
  month        = may,
  day          = {28},
  note         = {Personal communication via phone interview},
  howpublished = {Interview with the Stackable team, conducted by the author}
}

@inproceedings{araujo2022advancing,
  title={Advancing data architectures with data mesh implementations},
  author={Ara{\'u}jo Machado, In{\^e}s and Costa, Carlos and Santos, Maribel Yasmina},
  booktitle={International Conference on Advanced Information Systems Engineering},
  pages={10--18},
  year={2022},
  organization={Springer}
}

@inproceedings{DBLP:conf/centeris/Machado0S21,
  author       = {In{\^{e}}s Ara{\'{u}}jo Machado and
                  Carlos Costa and
                  Maribel Yasmina Santos},
  editor       = {Maria Manuela Cruz{-}Cunha and
                  Ricardo Martinho and
                  Rui Rijo and
                  Dulce Domingos and
                  Emanuel Peres},
  title        = {Data Mesh: Concepts and Principles of a Paradigm Shift in Data Architectures},
  booktitle    = {{CENTERIS} 2021 - International Conference on ENTERprise Information Systems},
  series       = {Procedia Computer Science},
  volume       = {196},
  pages        = {263--271},
  publisher    = {Elsevier},
  year         = {2021},
  url          = {https://doi.org/10.1016/j.procs.2021.12.013},
  doi          = {10.1016/J.PROCS.2021.12.013},
  bibsource    = {dblp computer science bibliography, https://dblp.org}
}

@article{abbasi2016big,
  title={Big data research in information systems: Toward an inclusive research agenda},
  author={Abbasi, Ahmed and Sarker, Suprateek and Chiang, Roger HL},
  journal={Journal of the association for information systems},
  volume={17},
  number={2},
  pages={3},
  year={2016}
}

@INPROCEEDINGS{10278300,
  author={Ashraf, Abdulrahman and Hassan, Ahmed and Mahdi, Hani},
  booktitle={2023 International Mobile, Intelligent, and Ubiquitous Computing Conference (MIUCC)}, 
  title={Key Lessons from Microservices for Data Mesh Adoption}, 
  year={2023},
  volume={},
  number={},
  pages={1-8},
  doi={10.1109/MIUCC58832.2023.10278300}}

@inproceedings{sang2017simplifying,
  title={Simplifying big data analytics systems with a reference architecture},
  author={Sang, Go Muan and Xu, Lai and De Vrieze, Paul},
  booktitle={Collaboration in a Data-Rich World: 18th IFIP WG 5.5 Working Conference on Virtual Enterprises, PRO-VE 2017, Vicenza, Italy, September 18-20, 2017, Proceedings 18},
  pages={242--249},
  year={2017},
  organization={Springer}
}

@INPROCEEDINGS{7916249,
  author={Sang, Go Muan and Xu, Lai and de Vrieze, Paul},
  booktitle={2016 10th International Conference on Software, Knowledge, Information Management and Applications (SKIMA)}, 
  title={A reference architecture for big data systems}, 
  year={2016},
  volume={},
  number={},
  pages={370-375},
  doi={10.1109/SKIMA.2016.7916249}}

@article{bode2024towards,
  title={Towards Avoiding the Data Mess: Industry Insights from Data Mesh Implementations},
  author={Bode, Jan and K{\"u}hl, Niklas and Kreuzberger, Dominik and Holtmann, Carsten},
  journal={IEEE Access},
  year={2024},
  publisher={IEEE}
}

@inproceedings{butte2022enterprise,
  title={Enterprise data strategy: a decentralized data mesh approach},
  author={Butte, Vijay Kumar and Butte, Sujata},
  booktitle={2022 International Conference on Data Analytics for Business and Industry (ICDABI)},
  pages={62--66},
  year={2022},
  organization={IEEE}
}

@article{armbrust2010view,
  title={A view of cloud computing},
  author={Armbrust, Michael and Fox, Armando and Griffith, Rean and Joseph, Anthony D and Katz, Randy and Konwinski, Andy and Lee, Gunho and Patterson, David and Rabkin, Ariel and Stoica, Ion and others},
  journal={Communications of the ACM},
  volume={53},
  number={4},
  pages={50--58},
  year={2010},
  publisher={ACM New York, NY, USA}
}

@article{10.1145/96602.96604,
author = {Sheth, Amit P. and Larson, James A.},
title = {Federated database systems for managing distributed, heterogeneous, and autonomous databases},
year = {1990},
issue_date = {Sept. 1990},
publisher = {Association for Computing Machinery},
address = {New York, NY, USA},
volume = {22},
number = {3},
issn = {0360-0300},
url = {https://doi.org/10.1145/96602.96604},
doi = {10.1145/96602.96604},
journal = {ACM Comput. Surv.},
month = sep,
pages = {183–236},
numpages = {54}
}

@inproceedings{he2016deep,
  title={Deep residual learning for image recognition},
  author={He, Kaiming and Zhang, Xiangyu and Ren, Shaoqing and Sun, Jian},
  booktitle={Proceedings of the IEEE conference on computer vision and pattern recognition},
  pages={770--778},
  year={2016}
}

@article{simonyan2014very,
  title={Very deep convolutional networks for large-scale image recognition},
  author={Simonyan, Karen and Zisserman, Andrew},
  journal={arXiv preprint arXiv:1409.1556},
  year={2014}
}

@article{krizhevsky2012imagenet,
  title={Imagenet classification with deep convolutional neural networks},
  author={Krizhevsky, Alex and Sutskever, Ilya and Hinton, Geoffrey E},
  journal={Advances in neural information processing systems},
  volume={25},
  year={2012}
}

@book{golab2022data,
  title={Data stream management},
  author={Golab, Lukasz and Ozsu, M Tamer},
  year={2022},
  publisher={Springer Nature}
}

@inproceedings{fathy2019unified,
  title={A unified access to heterogeneous big data through ontology-based semantic integration},
  author={Fathy, Naglaa and Gad, Walaa and Badr, Nagwa},
  booktitle={2019 Ninth International Conference on Intelligent Computing and Information Systems (ICICIS)},
  pages={387--392},
  year={2019},
  organization={IEEE}
}

@book{dehghani2022data,
  author    = {Zhamak Dehghani},
  title     = {Data Mesh: Delivering Data-Driven Value at Scale},
  year      = {2022},
  publisher = {O'Reilly Media},
  address   = {Sebastopol, CA},
  isbn      = {978-1492092391},
  url       = {https://www.oreilly.com/library/view/data-mesh/9781492092346/}
}

@misc{Dehghani2019,
  author = {Zhamak Dehghani},
  title = {How to Move Beyond a Monolithic Data Lake to a Distributed Data Mesh},
  institution = {Thoughtworks},
  address = {Chicago, IL, USA},
  month = {May},
  year = {2019},
  url = {https://martinfowler.com/articles/data-monolith-to-mesh.html}
}

@INPROCEEDINGS{7364082,
  author={Kiran, Mariam and Murphy, Peter and Monga, Inder and Dugan, Jon and Baveja, Sartaj Singh},
  booktitle={2015 IEEE International Conference on Big Data (Big Data)}, 
  title={Lambda architecture for cost-effective batch and speed big data processing}, 
  year={2015},
  volume={},
  number={},
  pages={2785-2792},
  doi={10.1109/BigData.2015.7364082}}

@article{martinez2024candil,
	title={CANDIL: A federated data fabric for network analytics},
	author={Martinez-Casanueva, Ignacio D and Bellido, Luis and Gonz{\'a}lez-S{\'a}nchez, Daniel and Lopez, Diego},
	journal={Future Generation Computer Systems},
	volume={158},
	pages={98--109},
	year={2024},
	publisher={Elsevier}
}

@article{davoudian2018survey,
	title={A survey on NoSQL stores},
	author={Davoudian, Ali and Chen, Liu and Liu, Mengchi},
	journal={ACM Computing Surveys (CSUR)},
	volume={51},
	number={2},
	pages={1--43},
	year={2018},
	publisher={ACM New York, NY, USA}
}

@article{moon2021study,
	title={A study on a distributed data fabric-based platform in a multi-cloud environment},
	author={Moon, Seok-Jae and Kang, Seong-Beom and Park, Byung-Joon},
	journal={International Journal of Advanced Culture Technology},
	volume={9},
	number={3},
	pages={321--326},
	year={2021},
	publisher={The International Promotion Agency of Culture Technology}
}

@article{abu2023structural,
	title={Structural equation modeling for impact of Data Fabric Framework on business decision-making and risk management},
	author={Abu Rumman, Amani and Al-Abbadi, Lina},
	journal={Cogent Business \& Management},
	volume={10},
	number={2},
	pages={2215060},
	year={2023},
	publisher={Taylor \& Francis}
}

@article{blohm2024data,
	title={Data products, data mesh, and data fabric: New paradigm (s) for data and analytics?},
	author={Blohm, Ivo and Wortmann, Felix and Legner, Christine and K{\"o}bler, Felix},
	journal={Business \& Information Systems Engineering},
	pages={1--10},
	year={2024},
	publisher={Springer}
}

@article{alvord2022big,
	title={Big data fabric architecture: How big data and data management frameworks converge to bring a new generation of competitive advantage for enterprises},
	author={Alvord, Micah M and Lu, Fengyu and Du, Boyang and Chen, Chia-An},
	journal={Enterprise Architecture Professional Journal},
	year={2022}
}

@article{hechler2023data,
	title={Data Fabric and Data Mesh Approaches With AI},
	author={Hechler, YCWE and Weihrauch, Maryela and Wu, Yan Catherine},
	journal={Berkeley, CA, USA: Apress Berkeley},
	year={2023},
	publisher={Springer}
}

@inproceedings{li2022distributed,
	title={A distributed data fabric architecture based on metadate knowledge graph},
	author={Li, XinChi and Yang, Mingchuan and Xia, Xiaoqing and Zhang, Kaicheng and Liu, Kang},
	booktitle={2022 5th International Conference on Data Science and Information Technology (DSIT)},
	pages={1--7},
	year={2022},
	organization={IEEE}
}

@INPROCEEDINGS{9671862,
	author={Priebe, Torsten and Neumaier, Sebastian and Markus, Stefan},
	booktitle={2021 IEEE International Conference on Big Data (Big Data)}, 
	title={Finding Your Way Through the Jungle of Big Data Architectures}, 
	year={2021},
	volume={},
	number={},
	pages={5994-5996},
	doi={10.1109/BigData52589.2021.9671862}
	}

@article{rieyan2024advanced,
	title={An advanced data fabric architecture leveraging homomorphic encryption and federated learning},
	author={Rieyan, Sakib Anwar and News, Md Raisul Kabir and Rahman, ABM Muntasir and Khan, Sadia Afrin and Zaarif, Sultan Tasneem Jawad and Alam, Md Golam Rabiul and Hassan, Mohammad Mehedi and Ianni, Michele and Fortino, Giancarlo},
	journal={Information Fusion},
	volume={102},
	pages={102004},
	year={2024},
	publisher={Elsevier}
}

@incollection{debellis2023interoperability,
	title={Interoperability Frameworks: Data Fabric and Data Mesh Architectures},
	author={DeBellis, Michael and Pinera, Livia and Connor, Christopher},
	booktitle={Data Science with Semantic Technologies},
	pages={267--286},
	year={2023},
	publisher={CRC Press}
}

@online{cdata2024,
	author       = {CData Software},
	title        = {Data Fabric vs. Data Virtualization},
	year         = {2024},
	url          = {https://www.cdata.com/blog/data-fabric-vs-data-virtualization},
	note         = {Accessed: 2024-11-22}
}

@article{macias2024data,
	title={Data fabric and digital twins: An integrated approach for data fusion design and evaluation of pervasive systems},
	author={Mac{\'\i}as, Aurora and Mu{\~n}oz, David and Navarro, Elena and Gonz{\'a}lez, Pascual},
	journal={Information Fusion},
	volume={103},
	pages={102-139},
	year={2024},
	publisher={Elsevier}
}

@article{akermi2023data,
	title={Data Virtualization Enabling Distributed Data Architectures: Data Fabric and Data Mesh},
	author={AKERMI, Montasser and TAIEB, Mohamed Ali HADJ and AOUICHA, Mohamed BEN},
	journal={International Journal of Computer Information Systems and Industrial Management Applications},
	volume={15},
	pages={12--12},
	year={2023}
}

@inproceedings{kuftinova2022data,
	title={Data fabric as an effective method of data management in traffic and road systems},
	author={Kuftinova, NG and Maksimychev, OI and Ostroukh, AV and Volosova, AV and Matukhina, EN},
	booktitle={2022 Systems of Signals Generating and Processing in the Field of on Board Communications},
	pages={1--4},
	year={2022},
	organization={IEEE}
}

@book{serra2024deciphering,
	title={Deciphering Data Architectures},
	author={Serra, James},
	year={2024},
	publisher={O'Reilly Media, Inc.}
}

@misc{gartner_fabric,
	author       = {Thanaraj, Robert and Beyer, Mark and Zaidi, Ehtisham},
	title        = {What Is Data Fabric Design?},
	year         = {2021},
	howpublished = {\url{https://info.cambridgesemantics.com/hubfs/What_Is_Data_Fabric_Gartner.pdf}},
	note         = {Accessed on 22.11.2024}
}

@article{zaidi2019data,
	title={Data fabrics add augmented intelligence to modernize your data integration},
	author={Zaidi, Ehitsham and Thoo, E and De Simoni, G and Beyer, M},
	journal={With Eric Thoo. Gartner Grou},
	volume={17},
	year={2019}
}

@incollection{otto2022evolution,
	title={The evolution of data spaces},
	author={Otto, Boris},
	booktitle={Designing data spaces: The ecosystem approach to competitive advantage},
	pages={3--15},
	year={2022},
	publisher={Springer International Publishing Cham}
}

@article{bacco2024data,
  title={What are data spaces? Systematic survey and future outlook},
  author={Bacco, Manlio and Kocian, Alexander and Chessa, Stefano and Crivello, Antonino and Barsocchi, Paolo},
  journal={Data in Brief},
  volume={57},
  pages={110969},
  year={2024},
  publisher={Elsevier}
}

@article{soininen2025data,
  title={What is a Data Space--Logical Architecture Model},
  author={Soininen, Juha-Pekka and Laatikainen, Gabriella},
  journal={Data in Brief},
  pages={111575},
  year={2025},
  publisher={Elsevier}
}

@article{putrama2024heterogeneous,
  title={Heterogeneous data integration: Challenges and opportunities},
  author={Putrama, I Made and Martinek, P{\'e}ter},
  journal={Data in Brief},
  volume={56},
  pages={110853},
  year={2024},
  publisher={Elsevier}
}

@article{tzanetos2024introduction,
  title={Introduction to the special issue “Data in Management and Decision Engineering”},
  author={Tzanetos, Alexandros and Dounias, Georgios},
  journal={Data in Brief},
  volume={55},
  pages={110711},
  year={2024}
}

@article{gessler2025business,
  title={Business models and organizational roles of data spaces: A framework for sustainable value creation},
  author={Gessler, Jens and Cencic, Maiara Rosa and Metzner, Christian and Wieker, Horst and Lindow, Kai and Schulz, Wolfgang H},
  journal={Data in Brief},
  pages={111795},
  year={2025},
  publisher={Elsevier}
}

@article{yun2025eicopilot,
  title={EICopilot: Search and Explore Enterprise Information over Large-scale Knowledge Graphs with LLM-driven Agents},
  author={Yun, Yuhui and Ye, Huilong and Li, Xinru and Li, Ruojia and Deng, Jingfeng and Li, Li and Xiong, Haoyi},
  journal={arXiv preprint arXiv:2501.13746},
  year={2025}
}

@techreport{plattform_industrie_2019_gaiax,
	author       = {{Plattform Industrie 4.0}},
	title        = {Project GAIA-X: A Federated Data Infrastructure as the Cradle of a Vibrant European Ecosystem},
	institution  = {Federal Ministry for Economic Affairs and Energy (BMWi)},
	address      = {D-11019 Berlin, Germany},
	year         = {2019},
	month        = {Oct.}
}

@article{lin2017lambda,
  title={The lambda and the kappa},
  author={Lin, Jimmy},
  journal={IEEE Internet Computing},
  volume={21},
  number={05},
  pages={60--66},
  year={2017},
  publisher={IEEE Computer Society}
}

@article{venugopal2006taxonomy,
  title={A taxonomy of data grids for distributed data sharing, management, and processing},
  author={Venugopal, Srikumar and Buyya, Rajkumar and Ramamohanarao, Kotagiri},
  journal={ACM Computing Surveys (CSUR)},
  volume={38},
  number={1},
  pages={3--es},
  year={2006},
  publisher={ACM New York, NY, USA}
}

@book{inmon2014data,
  title={Data architecture: a primer for the data scientist: big data, data warehouse and data vault},
  author={Inmon, William H and Linstedt, Daniel},
  year={2014},
  publisher={Morgan Kaufmann}
}

@Article{JJQV99IS,
  Title                    = {{Architecture and Quality in Data Warehouses: An Extended Repository Approach}},
  Author                   = {Matthias Jarke and Manfred A. Jeusfeld and C. Quix and Panos Vassiliadis},
  Journal                  = {Information Systems},
  Year                     = {1999},
  Number                   = {3},
  Pages                    = {229-253},
  Volume                   = {24}
}

@misc{mami2018teach,
	author = {Mami, Mohamed Nadjib and Jabeen, Hajira and Auer, Sören},
	year = {2018},
	title = {“Teach me to fish”: Querying Semantic Data Lakes},
	howpublished = {\url{https://www.researchgate.net/publication/322526357_'Teach_me_to_fish'_Querying_Semantic_Data_Lakes}},
	note = {Accessed: 2024-11-04}
}

@inproceedings{huai2014major,
	title={Major technical advancements in apache hive},
	author={Huai, Yin and Chauhan, Ashutosh and Gates, Alan and Hagleitner, Gunther and Hanson, Eric N and O'Malley, Owen and Pandey, Jitendra and Yuan, Yuan and Lee, Rubao and Zhang, Xiaodong},
	booktitle={Proceedings of the 2014 ACM SIGMOD international conference on Management of data},
	pages={1235--1246},
	year={2014}
}

@article{chaudhuri1997overview,
	title={An overview of data warehousing and OLAP technology},
	author={Chaudhuri, Surajit and Dayal, Umeshwar},
	journal={ACM Sigmod record},
	volume={26},
	number={1},
	pages={65--74},
	year={1997},
	publisher={ACM New York, NY, USA}
}

@article{bose2009advanced,
	title={Advanced analytics: opportunities and challenges},
	author={Bose, Ranjit},
	journal={Industrial Management \& Data Systems},
	volume={109},
	number={2},
	pages={155--172},
	year={2009},
	publisher={Emerald Group Publishing Limited}
}

@phdthesis{quix2003metadatenverwaltung,
	title={Metadatenverwaltung zur qualit{\"a}tsorientierten Informationslogistik in Data-Warehouse-Systemen},
	author={Quix, Christoph Josef},
	year={2003},
	school={Bibliothek der RWTH Aachen}
}

@book{kimball2013data,
	title={The data warehouse toolkit: The definitive guide to dimensional modeling},
	author={Kimball, Ralph and Ross, Margy},
	year={2013},
	publisher={John Wiley \& Sons}
}

@article{schneider2024lakehouse,
	title={The Lakehouse: State of the Art on Concepts and Technologies},
	author={Schneider, Jan and Gr{\"o}ger, Christoph and Lutsch, Arnold and Schwarz, Holger and Mitschang, Bernhard},
	journal={SN Computer Science},
	volume={5},
	number={5},
	pages={1--39},
	year={2024},
	publisher={Springer}
}

@article{harby2024data,
	title={Data Lakehouse: A survey and experimental study},
	author={Harby, Ahmed A and Zulkernine, Farhana},
	journal={Information Systems},
	pages={102460},
	year={2024},
	publisher={Elsevier}
}

@article{paton2023dataset,
	title={Dataset discovery and exploration: A survey},
	author={Paton, Norman W and Chen, Jiaoyan and Wu, Zhenyu},
	journal={ACM Computing Surveys},
	volume={56},
	number={4},
	pages={1--37},
	year={2023},
	publisher={ACM New York, NY, USA}
}

@article{ouellette2021ronin,
	title={RONIN: data lake exploration},
	author={Ouellette, Paul and Sciortino, Aidan and Nargesian, Fatemeh and Bashardoost, Bahar Ghadiri and Zhu, Erkang and Pu, Ken Q and Miller, Ren{\'e}e J},
	journal={Proceedings of the VLDB Endowment},
	volume={14},
	number={12},
	year={2021}
}

@INPROCEEDINGS{10020719,
	author={Harby, Ahmed A. and Zulkernine, Farhana},
	booktitle={2022 IEEE International Conference on Big Data (Big Data)}, 
	title={From Data Warehouse to Lakehouse: A Comparative Review}, 
	year={2022},
	volume={},
	number={},
	pages={389-395},
	doi={10.1109/BigData55660.2022.10020719}
}

@article{groger2021there,
	title={There is no AI without data},
	author={Gr{\"o}ger, Christoph},
	journal={Communications of the ACM},
	volume={64},
	number={11},
	pages={98--108},
	year={2021},
	publisher={ACM New York, NY, USA}
}

@inproceedings{ravat2019data,
	title={Data lakes: Trends and perspectives},
	author={Ravat, Franck and Zhao, Yan},
	booktitle={Database and Expert Systems Applications: 30th International Conference, DEXA 2019, Linz, Austria, August 26--29, 2019, Proceedings, Part I 30},
	pages={304--313},
	year={2019},
	organization={Springer}
}

@book{gorelik2016enterprise,
	author    = {Gorelik, Alex},
	title     = {The Enterprise Big Data Lake},
	year      = {2016},
	publisher = {O'Reilly Media},
	address   = {Sebastopol, CA, USA},
}

@book{zikopoulos2015bigdata,
	author    = {Zikopoulos, Paul and DeRoos, Dirk and Bienko, Chris and Buglio, Robert and Andrews, Mike},
	title     = {Big Data Beyond the Hype},
	edition   = {1st},
	year      = {2015},
	publisher = {McGraw-Hill Education},
	address   = {New York, NY, USA},
}

@book{madsen2015enterprisedatalake,
	author    = {Madsen, Mark},
	title     = {How to Build an Enterprise Data Lake: Important Considerations before Jumping},
	year      = {2015},
	publisher = {Third Nature Inc.},
	address   = {San Mateo, CA, USA},
}

@article{patel2017datalakegovernance,
	author    = {Patel, Pratik and Wood, Gordon and Diaz, Alberto},
	title     = {Data Lake Governance Best Practices},
	journal   = {Dzone Guide Big Data—Data Sci. Adv. Anal.},
	volume    = {4},
	pages     = {6--7},
	year      = {2017},
}

@book{sharma2018architectingdatalakes,
	author    = {Sharma, Ben},
	title     = {Architecting Data Lakes—Data Management Architectures for Advanced Business Use Cases},
	edition   = {2nd},
	year      = {2018},
	publisher = {O'Reilly Media},
	address   = {Sebastopol, CA, USA},
}

@inproceedings{10.1145/3209281.3209335,
author = {Oliveira, Marcelo Iury S. and L\'{o}scio, Bernadette Farias},
title = {What is a data ecosystem?},
year = {2018},
isbn = {9781450365260},
publisher = {ACM},
address = {New York, NY, USA},
url = {https://doi.org/10.1145/3209281.3209335},
doi = {10.1145/3209281.3209335},
booktitle = {Proceedings of the 19th Annual International Conference on Digital Government Research: Governance in the Data Age},
articleno = {74},
numpages = {9},
location = {Delft, The Netherlands},
series = {dg.o '18}
}

@inproceedings{giebler2021data,
  title={The data lake architecture framework},
  author={Giebler, Corinna and Gr{\"o}ger, Christoph and Hoos, Eva and Eichler, Rebecca and Schwarz, Holger and Mitschang, Bernhard},
  booktitle={BTW 2021},
  pages={351--370},
  year={2021},
  organization={Gesellschaft f{\"u}r Informatik, Bonn}
}

@article{muller2008reference,
	title={A reference architecture primer},
	author={Muller, Gerrit},
	journal={Eindhoven Univ. of Techn., Eindhoven, White paper},
	pages={24},
	year={2008}
}

@article{azzabi2024data,
	title={Data Lakes: A Survey of Concepts and Architectures},
	author={Azzabi, Sarah and Alfughi, Zakiya and Ouda, Abdelkader},
	journal={Computers},
	volume={13},
	number={7},
	pages={183},
	year={2024},
	publisher={MDPI}
}

@article{nambiar2022overview,
	title={An overview of data warehouse and data lake in modern enterprise data management},
	author={Nambiar, Athira and Mundra, Divyansh},
	journal={Big data and cognitive computing},
	volume={6},
	number={4},
	pages={132},
	year={2022},
	publisher={MDPI}
}

@misc{dixon2010pentaho,
	author = {Dixon, James},
	title = {Pentaho, Hadoop, and Data Lakes | James Dixon’s Blog},
	year = {2010},
	url = {https://jamesdixon.wordpress.com/2010/10/14/pentaho-hadoop-and-data-lakes/},
	note = {[Online; accessed 22-Oct-2024]}
}

@misc{gartner2014data,
	author = {Gartner, Inc.},
	title = {Gartner Says Beware of the Data Lake Fallacy},
	year = {2014},
	url = {https://www.gartner.com/en/newsroom/press-releases/2014-07-28-gartner-says-beware-of-the-data-lake-fallacy},
	note = {[Online; accessed 29-Jun-2025]}
}

@misc{dixon2014revisited,
	author = {Dixon, James},
	title = {Data Lakes Revisited},
	year = {2014},
	url = {https://jamesdixon.wordpress.com/2014/09/25/data-lakes-revisited/},
	note = {[Online; accessed 22-Oct-2024]}
}

@book{inmon2005building,
	title={Building the data warehouse},
	author={Inmon, William H},
	year={2005},
	publisher={John wiley \& sons}
}

@inproceedings{DBLP:conf/ijcai/XiaoCKLPRZ18,
  author       = {Guohui Xiao and
                  Diego Calvanese and
                  Roman Kontchakov and
                  Domenico Lembo and
                  Antonella Poggi and
                  Riccardo Rosati and
                  Michael Zakharyaschev},
  editor       = {J{\'{e}}r{\^{o}}me Lang},
  title        = {Ontology-Based Data Access: {A} Survey},
  booktitle    = {Proceedings of the Twenty-Seventh International Joint Conference on
                  Artificial Intelligence, {IJCAI} 2018, July 13-19, 2018, Stockholm,
                  Sweden},
  pages        = {5511--5519},
  publisher    = {ijcai.org},
  year         = {2018},
  url          = {https://doi.org/10.24963/ijcai.2018/777},
  doi          = {10.24963/IJCAI.2018/777},
  bibsource    = {dblp computer science bibliography, https://dblp.org}
}

@incollection{alla2020mlops,
  title={What is mlops?},
  author={Alla, Sridhar and Adari, Suman Kalyan},
  booktitle={Beginning MLOps with MLFlow: Deploy Models in AWS SageMaker, Google Cloud, and Microsoft Azure},
  pages={79--124},
  year={2020},
  publisher={Springer}
}

@article{moniruzzaman2013nosql,
  title={Nosql database: New era of databases for big data analytics-classification, characteristics and comparison},
  author={Moniruzzaman, ABM and Hossain, Syed Akhter},
  journal={arXiv preprint arXiv:1307.0191},
  year={2013}
}

@article{DBLP:journals/internet/DeckerMHFKBEH00,
	author       = {Stefan Decker and
	Sergey Melnik and
	Frank van Harmelen and
	Dieter Fensel and
	Michel C. A. Klein and
	Jeen Broekstra and
	Michael Erdmann and
	Ian Horrocks},
	title        = {The Semantic Web: The Roles of {XML} and {RDF}},
	journal      = {{IEEE} Internet Comput.},
	volume       = {4},
	number       = {5},
	pages        = {63--74},
	year         = {2000},
	url          = {https://doi.org/10.1109/4236.877487},
	doi          = {10.1109/4236.877487},
	bibsource    = {dblp computer science bibliography, https://dblp.org}
}

@book{linstedt2015building,
  title={Building a scalable data warehouse with data vault 2.0},
  author={Linstedt, Daniel and Olschimke, Michael},
  year={2015},
  publisher={Morgan Kaufmann}
}

@Book{DWQBuch,
  Title                    = {Fundamentals of Data Warehouses},
  Editor                   = {Matthias Jarke and Maurizio Lenzerini and Yannis Vassiliou and Panos Vassiliadis},
  Publisher                = {Springer-Verlag},
  Year                     = {2003},
  Edition                  = {2},
  Booktitle                = {Fundamentals of Data Warehouses}
}

@ARTICLE{hai2023data,
	author={Hai, Rihan and Koutras, Christos and Quix, Christoph and Jarke, Matthias},
	journal={IEEE TKDE}, 
	title={Data Lakes: A Survey of Functions and Systems}, 
	year={2023},
	volume={35},
	number={12},
	pages={12571-12590},
	temp_doi={10.1109/TKDE.2023.3270101}
}

@article{DBLP:journals/dke/EichlerGGSM21,
	author       = {Rebecca Eichler and
	Corinna Giebler and
	Christoph Gr{\"{o}}ger and
	Holger Schwarz and
	Bernhard Mitschang},
	title        = {Modeling metadata in data lakes - {A} generic model},
	journal      = {Data Knowl. Eng.},
	volume       = {136},
	pages        = {101931},
	year         = {2021},
	url          = {https://doi.org/10.1016/j.datak.2021.101931},
	doi          = {10.1016/J.DATAK.2021.101931},
	bibsource    = {dblp computer science bibliography, https://dblp.org}
}

@book{international2017dama,
	title={DAMA-DMBOK: Data management body of knowledge},
	author={Data Administration Management Association and others},
	year={2017},
	publisher={Technics Publications, LLC}
}

@inproceedings{badampudi2015experiences,
	title={Experiences from using snowballing and database searches in systematic literature studies},
	author={Badampudi, Deepika and Wohlin, Claes and Petersen, Kai},
	booktitle={Proceedings of the 19th international conference on evaluation and assessment in software engineering},
	pages={1--10},
	year={2015}
}

@inproceedings{armbrust2021lakehouse,
	title={Lakehouse: a new generation of open platforms that unify data warehousing and advanced analytics},
	author={Armbrust, Michael and Ghodsi, Ali and Xin, Reynold and Zaharia, Matei},
	booktitle={Proceedings of CIDR},
	volume={8},
	pages={28},
	year={2021}
}

@article{sawadogo2021data,
	title={On data lake architectures and metadata management},
	author={Sawadogo, Pegdwend{\'e} and Darmont, J{\'e}r{\^o}me},
	journal={JJIS},
	year={2021},
}

@InProceedings{farid2016clams:short,
	author       = {Farid, Mina and Roatis, Alexandra and Ilyas, Ihab F and Hoffmann, Hella-Franziska and Chu, Xu},
	booktitle    = {Proc.\ {SIGMOD}},
	title        = {CLAMS: Bringing Quality to Data Lakes},
	year         = {2016},
}

@article{scholly2021coining,
	title={Coining goldMEDAL: a new contribution to data lake generic metadata modeling},
	author={Scholly, Etienne and Sawadogo, Pegdwend{\'e} and Liu, Pengfei and Espinosa-Oviedo, Javier Alfonso and Favre, C{\'e}cile and Loudcher, Sabine and Darmont, J{\'e}r{\^o}me and No{\^u}s, Camille},
	journal={arXiv preprint arXiv:2103.13155},
	year={2021}
}

@inproceedings{10.1145/3340531.3417426,
	author = {Galhotra, Sainyam and Khurana, Udayan},
	title = {Semantic Search over Structured Data},
	year = {2020},
	isbn = {9781450368599},
	publisher = {Association for Computing Machinery},
	address = {New York, NY, USA},
	doi = {10.1145/3340531.3417426},
	booktitle = {Proceedings of the 29th ACM International Conference on Information \& Knowledge Management},
	pages = {3381–3384},
	numpages = {4},
	location = {Virtual Event, Ireland},
}

@article{pomp2018applying,
	title={Applying semantics to reduce the time to analytics within complex heterogeneous infrastructures},
	author={Pomp, Andr{\'e} and Paulus, Alexander and Kirmse, Andreas and Kraus, Vadim and Meisen, Tobias},
	journal={Technologies},
	volume={6},
	number={3},
	pages={86},
	year={2018},
	publisher={MDPI}
}

@Article{Quix2016,
	author    = {Christoph Quix and Rihan Hai and Ivan Vatov},
	journal   = {Complex Syst. Informatics Model. Q.},
	title     = {Metadata Extraction and Management in Data Lakes With {GEMMS}},
	year      = {2016},
	bibsource = {dblp computer science bibliography, https://dblp.org},
	temp_doi= {10.7250/csimq.2016-9.04},
}

@inproceedings{sawadogo2019metadata,
	title={Metadata systems for data lakes: models and features},
	author={Sawadogo, Pegdwend{\'e} N and Scholly, Etienne and Favre, C{\'e}cile and Ferey, Eric and Loudcher, Sabine and Darmont, J{\'e}r{\^o}me},
	booktitle={European conference on advances in databases and information systems},
	pages={440--451},
	year={2019},
	organization={Springer}
}

@inproceedings{eichler2020handle,
	title={Handle-a generic metadata model for data lakes},
	author={Eichler, Rebecca and Giebler, Corinna and Gr{\"o}ger, Christoph and Schwarz, Holger and Mitschang, Bernhard},
	booktitle={International Conference on Big Data Analytics and Knowledge Discovery},
	pages={73--88},
	year={2020},
	organization={Springer}
}

@inproceedings{hoseini2023sedar,
	title={SEDAR: A Semantic Data Reservoir for Heterogeneous Datasets},
	author={Hoseini, Sayed and Ali, Ahmed and Shaker, Haron and Quix, Christoph},
	booktitle={Proceedings of the 32nd ACM International Conference on Information and Knowledge Management},
	pages={5056--5060},
	year={2023}
}

@article{van2003data,
  title={Data Architecture: Blueprints for Data.},
  author={Van den Hoven, John},
  journal={Information systems management},
  volume={20},
  number={1},
  year={2003},
  publisher={Taylor \& Francis}
}

@book{date1977introduction,
  title={An introduction to database systems},
  author={Date, Christopher John},
  year={1977},
  publisher={Pearson Education India}
}

@article{geisler2021knowledge,
  title={Knowledge-driven data ecosystems toward data transparency},
  author={Geisler, Sandra and Vidal, Maria-Esther and Cappiello, Cinzia and L{\'o}scio, Bernadette Farias and Gal, Avigdor and Jarke, Matthias and Lenzerini, Maurizio and Missier, Paolo and Otto, Boris and Paja, Elda and others},
  journal={ACM Journal of Data and Information Quality (JDIQ)},
  volume={14},
  number={1},
  pages={1--12},
  year={2021},
  publisher={ACM New York, NY}
}

@inproceedings{DBLP:conf/birthday/JarkeQ17,
  author       = {Matthias Jarke and
                  Christoph Quix},
  editor       = {Jordi Cabot and
                  Cristina G{\'{o}}mez and
                  Oscar Pastor and
                  Maria{-}Ribera Sancho and
                  Ernest Teniente},
  title        = {On Warehouses, Lakes, and Spaces: The Changing Role of Conceptual
                  Modeling for Data Integration},
  booktitle    = {Conceptual Modeling Perspectives},
  pages        = {231--245},
  publisher    = {Springer},
  year         = {2017},
  url          = {https://doi.org/10.1007/978-3-319-67271-7\_16},
  doi          = {10.1007/978-3-319-67271-7\_16},
  bibsource    = {dblp computer science bibliography, https://dblp.org}
}

@InProceedings{Endris2019,
	author    = {Kemele M. Endris and Philipp D. Rohde and Maria{-}Esther Vidal and S{\"{o}}ren Auer},
	booktitle = {Proc.\ {DEXA}},
	title     = {Ontario: Federated Query Processing Against a Semantic Data Lake},
	year      = {2019},
	pages     = {379--395},
	publisher = {Springer},
	series    = {LNCS},
	volume    = {11706},
	bibsource = {dblp computer science bibliography, https://dblp.org},
	doi       = {10.1007/978-3-030-27615-7      \_{2}{9}},
}

@InProceedings{Bagozi2019,
	author    = {Ada Bagozi and Devis Bianchini and Valeria De Antonellis and Massimiliano Garda and Michele Melchiori},
	booktitle = {Proc.\ {OTM} Conf.},
	title     = {Personalised Exploration Graphs on Semantic Data Lakes},
	year      = {2019},
	pages     = {22--39},
	publisher = {Springer},
	series    = {LNCS},
	volume    = {11877},
	bibsource = {dblp computer science bibliography, https://dblp.org},
	doi       = {10.1007/978-3-030-33246-4          \_2},
}

@inproceedings{DBLP:conf/semweb/MamiGSJA019a,
	author       = {Mohamed Nadjib Mami and
	Damien Graux and
	Simon Scerri and
	Hajira Jabeen and
	S{\"{o}}ren Auer and
	Jens Lehmann},
	editor       = {Chiara Ghidini and
	Olaf Hartig and
	Maria Maleshkova and
	Vojtech Sv{\'{a}}tek and
	Isabel F. Cruz and
	Aidan Hogan and
	Jie Song and
	Maxime Lefran{\c{c}}ois and
	Fabien Gandon},
	title        = {Squerall: Virtual Ontology-Based Access to Heterogeneous and Large
	Data Sources},
	booktitle    = {The Semantic Web - {ISWC} 2019 - 18th International Semantic Web Conference,
	Auckland, New Zealand, October 26-30, 2019, Proceedings, Part {II}},
	series       = {Lecture Notes in Computer Science},
	volume       = {11779},
	pages        = {229--245},
	publisher    = {Springer},
	year         = {2019},
	temp_doi= {10.1007/978-3-030-30796-7\_15},
	bibsource    = {dblp computer science bibliography, https://dblp.org}
}

@inproceedings{hoseini2024enhancing,
	title={Enhancing Machine Learning Capabilities in Data Lakes with AutoML and LLMs},
	author={Hoseini, Sayed and Ibbels, Maximilian and Quix, Christoph},
	booktitle={European Conference on Advances in Databases and Information Systems},
	pages={184--198},
	year={2024},
	organization={Springer}
}

@Article{DBLP:journals/debu/HalevyKNOPRW16,
	author    = {Alon Y. Halevy and Flip Korn and Natalya Fridman Noy and Christopher Olston and Neoklis Polyzotis and Sudip Roy and Steven Euijong Whang},
	journal   = {{IEEE} Data Eng. Bull.},
	title     = {Managing Google's data lake: an overview of the Goods system},
	year      = {2016},
	bibsource = {dblp computer science bibliography, https://dblp.org},
	url       = {http://sites.computer.org/debull/A16sept/p5.pdf},
}

@inproceedings{ramakrishnan2017azure,
	title={Azure data lake store: a hyperscale distributed file service for big data analytics},
	author={Ramakrishnan, Raghu and Sridharan, Baskar and Douceur, John R and Kasturi, Pavan and Krishnamachari-Sampath, Balaji and Krishnamoorthy, Karthick and Li, Peng and Manu, Mitica and Michaylov, Spiro and Ramos, Rog{\'e}rio and others},
	booktitle={Proceedings of the 2017 ACM International Conference on Management of Data},
	pages={51--63},
	year={2017}
}

@inproceedings{Dibowski2020UsingST,
	title={Using Semantic Technologies to Manage a Data Lake: Data Catalog, Provenance and Access Control},
	author={Henrik Dibowski and Stefan Schmid and Yulia Svetashova and Cory Andrew Henson and Tuan Tran},
	booktitle={SSWS@ISWC},
	year={2020},
	url={https://api.semanticscholar.org/CorpusID:229357020}
}

@article{zhao2021data,
	title={Data Lake Ingestion Management},
	author={Zhao, Yan and Megdiche, Imen and Ravat, Franck},
	journal={arXiv preprint arXiv:2107.02885},
	year={2021}
}

@InProceedings{Dibowski2020a,
	author    = {Henrik Dibowski and Stefan Schmid and Yulia Svetashova and Cory Henson and Tuan Tran},
	booktitle = {Proc.\ Scalable Semantic Web Knowledge Base Systems Workshop},
	title     = {Using Semantic Technologies to Manage a Data Lake: Data Catalog, Provenance and Access Control},
	year      = {2020},
	pages     = {65--80},
	series = {CEUR WS},
	volume    = {2757},
	bibsource = {dblp computer science bibliography, https://dblp.org},
	temp_url = {http://ceur-ws.org/Vol-2757/SSWS2020_paper5.pdf},
}

@inproceedings{Schneider2023,
	author = {Schneider, Jan and Gr{\"{o}}ger, Christoph and Lutsch, Arnold and
	Schwarz, Holger and Mitschang, Bernhard},
	booktitle = {Proceedings of the 25th International Conference on Enterprise
	Information Systems (ICEIS 2023)},
	publisher = {Scitepress 2023},
	title = {Assessing the Lakehouse: Analysis, Requirements and Definition},
	year = {2023}
}

@InProceedings{Pomp2017,
	author    = {Andr{\'{e}} Pomp and Alexander Paulus and Sabina Jeschke and Tobias Meisen},
	booktitle = {Proc.\ {ICEIS}},
	title     = {{ESKAPE:} Information Platform for Enabling Semantic Data Processing},
	year      = {2017},
	pages     = {644--655},
	publisher = {SciTePress},
	bibsource = {dblp computer science bibliography, https://dblp.org},
	doi       = {10.5220/0006324906440655},
}

@article{armbrust2020delta,
	title={Delta lake: high-performance ACID table storage over cloud object stores},
	author={Armbrust, Michael and Das, Tathagata and Sun, Liwen and Yavuz, Burak and Zhu, Shixiong and Murthy, Mukul and Torres, Joseph and van Hovell, Herman and Ionescu, Adrian and {\L}uszczak, Alicja and others},
	journal={Proceedings of the VLDB Endowment},
	volume={13},
	number={12},
	pages={3411--3424},
	year={2020},
	publisher={VLDB Endowment}
}

@article{dibowski2021using,
	title={Using knowledge graphs to manage a data lake},
	author={Dibowski, Henrik and Schmid, Stefan},
	year={2021},
	publisher={Gesellschaft f{\"u}r Informatik, Bonn}
}

@inproceedings{DBLP:conf/semweb/KalayciGLXMKC20,
	author    = {Elem G{\"{u}}zel Kalayci and
	Irl{\'{a}}n Grangel{-}Gonz{\'{a}}lez and
	Felix L{\"{o}}sch and
	Guohui Xiao and
	Anees ul Mehdi and
	Evgeny Kharlamov and
	Diego Calvanese},
	title     = {Semantic Integration of Bosch Manufacturing Data Using Virtual Knowledge Graphs},
	booktitle = {Proc.\ {ISWC}},
	series    = {LNCS},
	volume    = {12507},
	pages     = {464--481},
	publisher = {Springer},
	year      = {2020},
	bibsource = {dblp computer science bibliography, https://dblp.org}
}

@article{laney20013d,
  title={3D data management: Controlling data volume, velocity and variety},
  author={Laney, Doug and others},
  journal={META group research note},
  volume={6},
  number={70},
  pages={1},
  year={2001},
  publisher={Stanford}
}

@InProceedings{Diamantini2018,
	author    = {Claudia Diamantini and Paolo Lo Giudice and Lorenzo Musarella and Domenico Potena and Emanuele Storti and Domenico Ursino},
	booktitle = {Proc.\ {ADBIS} Short Papers \& Workshops},
	title     = {A New Metadata Model to Uniformly Handle Heterogeneous Data Lake Sources},
	year      = {2018},
	pages     = {165--177},
	publisher = {Springer},
	series    = {CCIS},
	volume    = {909},
	bibsource = {dblp computer science bibliography, https://dblp.org},
	doi       = {10.1007/978-3-030-00063-9        \_{1}{7}},
}

@article{hoseini2023semantic,
	title = {A survey on semantic data management as intersection of ontology-based data access, semantic modeling and data lakes},
	journal = {Journal of Web Semantics},
	volume = {81},
	pages = {100819},
	year = {2024},
	issn = {1570-8268},
	doi = {https://doi.org/10.1016/j.websem.2024.100819},
	author = {Sayed Hoseini and Johannes Theissen-Lipp and Christoph Quix},
}

@inproceedings{hoseini2024coatings,
  title={Coatings Intelligence: Data-driven Automation for Chemistry 4.0},
  author={Hoseini, Sayed and Zhang, Gaoyuan and Polke, Dominik and Surjana, Alvin and Wagner, Lasse and Schmitz, Christian and Quix, Christoph},
  booktitle={2024 IEEE 7th International Conference on Industrial Cyber-Physical Systems (ICPS)},
  pages={1--6},
  year={2024},
  organization={IEEE}
}

@InProceedings{Calvanese2021,
	author    = {Diego Calvanese and Linfang Ding and Alessandro Mosca and Guohui Xiao},
	booktitle = {Proc.\ Italian {SIGCHI} (CHItaly)},
	title     = {Realizing Ontology-based Reusable Interfaces for Data Access via Virtual Knowledge Graphs},
	year      = {2021},
	pages     = {35:1--35:5},
	publisher = {{ACM}},
	bibsource = {dblp computer science bibliography, https://dblp.org},
	doi       = {10.1145/3464385.3464744},
}

@InProceedings{Paulus2021,
	author    = {Alexander Paulus and Andreas Burgdorf and Andr{\'{e}} Pomp and Tobias Meisen},
	booktitle = {Proc.\ 15th {IEEE} {ICSC}},
	title     = {Recent Advances and Future Challenges of Semantic Modeling},
	year      = {2021},
	pages     = {70--75},
	publisher = {{IEEE}},
	bibsource = {dblp computer science bibliography, https://dblp.org},
	doi       = {10.1109/ICSC50631.2021.00016},
}

@article{DBLP:journals/dbsk/HentschelDFHO25,
  author       = {Martin Hentschel and
                  Jonathan Dees and
                  Florian Funke and
                  Max Heimel and
                  Ismail Oukid},
  title        = {Building a Data Management System for the Cloud: Lessons Learned and
                  Future Directions},
  journal      = {Datenbank-Spektrum},
  volume       = {25},
  number       = {1},
  pages        = {17--28},
  year         = {2025},
  url          = {https://doi.org/10.1007/s13222-025-00494-9},
  doi          = {10.1007/S13222-025-00494-9},
  bibsource    = {dblp computer science bibliography, https://dblp.org}
}

@InProceedings{hai2016constance,
	author    = {Hai, Rihan and Geisler, Sandra and Quix, Christoph},
	booktitle = {Proc.\ {SIGMOD}},
	title     = {Constance: An intelligent data lake system},
	year      = {2016}
}

@article{10.14778/3352063.3352116,
	author = {Nargesian, Fatemeh and Zhu, Erkang and Miller, Ren\'{e}e J. and Pu, Ken Q. and Arocena, Patricia C.},
	title = {Data lake management: challenges and opportunities},
	year = {2019},
	issue_date = {August 2019},
	publisher = {VLDB Endowment},
	volume = {12},
	number = {12},
	issn = {2150-8097},
	temp_doi= {10.14778/3352063.3352116},
	journal = {Proc. VLDB Endow.},
	month = {aug},
	pages = {1986–1989},
	numpages = {4}
}

@InCollection{Quix2019,
	author    = {Christoph Quix and Rihan Hai},
	booktitle = {Encyclopedia of Big Data Technologies},
	publisher = {Springer},
	title     = {Data Lake},
	year      = {2019},
	bibsource = {dblp computer science bibliography, https://dblp.org},
	doi       = {10.1007/978-3-319-63962-8          \_{7}{-}{1}},
}

@inproceedings{Draining_the_Data_Swamp,
	author = {Brackenbury, Will and Liu, Rui and Mondal, Mainack and Elmore, Aaron J. and Ur, Blase and Chard, Kyle and Franklin, Michael J.},
	title = {Draining the Data Swamp: A Similarity-based Approach},
	year = {2018},
	isbn = {9781450358279},
	publisher = {Association for Computing Machinery},
	address = {New York, NY, USA},
	doi = {10.1145/3209900.3209911},
	booktitle = {Proceedings of the Workshop on Human-In-the-Loop Data Analytics},
	articleno = {13},
	numpages = {7},
	location = {Houston, TX, USA},
	series = {HILDA '18}
}

@article{DBLP:journals/pvldb/BeheshtiBNT18,
  author       = {Amin Beheshti and
                  Boualem Benatallah and
                  Reza Nouri and
                  Alireza Tabebordbar},
  title        = {CoreKG: a Knowledge Lake Service},
  journal      = {Proc. {VLDB} Endow.},
  volume       = {11},
  number       = {12},
  pages        = {1942--1945},
  year         = {2018},
  url          = {http://www.vldb.org/pvldb/vol11/p1942-beheshti.pdf},
  doi          = {10.14778/3229863.3236230},
  bibsource    = {dblp computer science bibliography, https://dblp.org}
}

@inproceedings{atzori2024dataspaces,
  title={Dataspaces: Concepts, architectures and initiatives},
  author={Atzori, Maurizio and Ciaramella, Angelo and Diamantini, Claudia and Martino, BD and Distefano, Salvatore and Facchinetti, Tullio and Montecchiani, Fabrizio and Nocera, Antonino and Ruffo, Giancarlo and Trasarti, Roberto and others},
  booktitle={CEUR workshop proceedings},
  volume={3606},
  year={2024},
  organization={CEUR-WS}
}

\end{document}